\documentclass[journal]{vgtc}                     

\onlineid{1005}

\vgtccategory{Research}

\vgtcpapertype{Theoretical \& Empirical}

\title{Visualization Vibes: \\ The Socio-Indexical Function of Visualization Design}

\author{%
  \authororcid{Michelle Morgenstern*}{0000-0002-6321-6613}
  \authororcid{Amy Rae Fox*}{0000-0003-0995-7899},
  \authororcid{Graham M. Jones}{0000-0001-6435-7066}, 
  \authororcid{Arvind Satyanarayan}{0000-0001-5564-635X}
}

\authorfooter{
    \item 
    \textbf{*} indicates equal contribution. Morgenstern \& Jones are affiliated with MIT Department of Anthropology. Fox \& Satyanarayan are affiliated with MIT CSAIL. Email: [mjmorgen, amyfox, gmj, arvindsatya] @mit.edu 
}

\abstract{%
  In contemporary information ecologies saturated with misinformation, disinformation, and a distrust of science itself, public data communication faces significant hurdles. Although visualization research has broadened criteria for effective design, governing paradigms privilege the accurate and efficient transmission of data. Drawing on theory from linguistic anthropology, we argue that such approaches\textemdash focused on encoding and decoding propositional content\textemdash cannot fully account for how people engage with visualizations and why particular visualizations might invite adversarial or receptive responses. In this paper, we present evidence that data visualizations communicate not only semantic, propositional meaning\textemdash meaning about data\textemdash but also social, \textit{indexical } meaning\textemdash meaning beyond data. From a series of ethnographically-informed interviews, we document how readers make rich and varied assessments of a visualization’s “vibes”\textemdash inferences about the social provenance of a visualization based on its design features. Furthermore, these social attributions have the power to influence reception, as readers’ decisions about how to engage with a visualization concern not only content, or even aesthetic appeal, but also their sense of alignment or disalignment with the entities they imagine to be involved in its production and circulation. We argue these inferences hinge on a function of human sign systems that has thus far been little studied in data visualization: \textit{socio-indexicality}, whereby the formal features (rather than the content) of communication evoke social contexts, identities, and characteristics. Demonstrating the presence and significance of this socio-indexical function in visualization, this paper offers both a conceptual foundation and practical intervention for troubleshooting breakdowns in public data communication.
}


\teaser{
  \includegraphics[width=\linewidth, 
    alt={Four images of data visualizations that have had their titles and labels replaced with placeholder text. The images range from a simple pie chart to an artistic rendering of bar charts reclining in hospital beds. Below each visualization is two quotations from interviewees about the visualization's "vibes"--the social contexts, personae, and characteristics it evokes. For example, the first visualization is described as having "Microsoft Office vibes" and the the second, third, and fourth as looking like "a New Yorker Cartoon", "someone's business presentation", and "something a Boomer put on Facebook", respectively.}
    ]{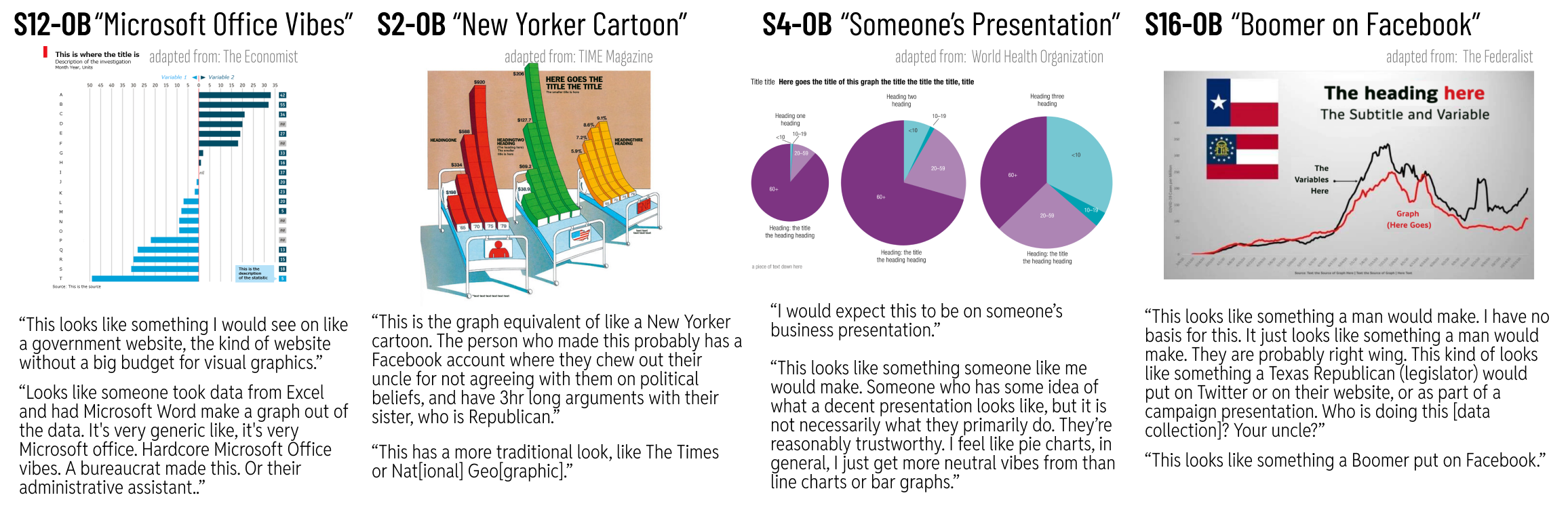}
  \caption{
  \textbf{Socio-Indexical Inferences from Message-Obscured Visualizations} 
  Four images of "message-obscured" visualizations, followed by examples of socio-indexical inferences made by participants in response to open-ended interview prompts.}
  \label{fig:teaser}
}

\graphicspath{{FIGURES/}{./}} 

\usepackage{mathptmx}                  

\usepackage{censor}

\usepackage{etoolbox}

\usepackage{csquotes} 

\usepackage{booktabs}
\usepackage{graphicx}

\usepackage{accessibility}

\usepackage{stfloats}

\renewenvironment{displayquote}
  { 
    \par
    \setlength{\leftskip}{0em}  
    \setlength{\rightskip}{0em} 
    \setlength{\parindent}{0pt}   
    \setlength{\parskip}{0.35\baselineskip}  
    \it
  }
  { 
    \par
    \vspace{0.5\baselineskip} 
  }

\begin{document}


\firstsection{Introduction}

\maketitle

\label{sec:introduction}

Communication breakdown characterizes much of our contemporary information ecology. Public data communication faces urgent challenges in reaching 
polarized media environments with adversarial audiences 
that increasingly disavow
the validity of science itself. At the same time, visualization (and visual media, more broadly) is being harnessed to propagate misinformation and disinformation. \cite{leeViralVisualizationsHow2021, murphyFakeNewsWeb2023a}. 
However, the governing paradigm of visualization \cite{CardCHI11999}\textemdash that visualizations function as a conduit for transmitting \emph{propositional information} or insights about data\textemdash offers inadequate explanations for these challenges. This paradigm predisposes researchers, designers, and readers alike to diagnose communication breakdowns as essentially a problem of the encoding-decoding process \cite{reddyConduitMetaphorCase1993}. Therefore, to combat failures in public data communication, we primarily look for solutions that intervene in that process: either developing design guidelines to more clearly and efficiently represent the data (the visualization’s propositional meaning), or improving data and visualization literacy to increase readers ability to quickly and accurately extract propositional information. 

While accurately communicating propositional meaning is undeniably important, a wealth of research in linguistic anthropology and sociolinguistics shows that natural language does more than just communicate semantic, propositional meaning (i.e. the content of the utterance, what it is about). A substantial body of evidence demonstrates that language also has a \textit{socio-indexical function}: any 
utterance also conveys ideas about the identities of, and therefore social relationships between, speakers and listeners, because 
formal features of language (e.g. aspects of accent, dialect, register, and choice of particular jargon or slang) point to\textemdash that is, \textit{index}\textemdash social categories, contexts, and characteristics. 
 
As a simple example, in spoken North American Englishes, the deletion of the 't' and 'd' sounds at the end of words (e.g., \emph{firs' chil'} instead of \emph{first child}) signals informality and, therefore, can mark a social situation as a relaxed encounter between peers.
Similarly, people often view popular styles of digital communication (so-called ``txt speak'') as indicating youthfulness and ignorance or 
laziness, despite evidence that these stylistic choices involve skill, creativity, and added effort \cite{jones2009enquoting, michelle-diss}. Drawing such social inferences, even when faulty, is not a failure of communication, but rather reflects a central feature of how language connects people as social beings, allowing them to perceive and navigate similarities and differences with others. 

Moreover, 
socio-indexical inferences can have profound implications 
for how propositional content is interpreted. For example, studies have demonstrated that when transcribing recordings of speech they infer to be "Black", court reporters in Philadelphia interpreted utterances as having underlying criminal meaning \cite{jones2019testifyingblack}. While socio-indexical meaning\textemdash and how it affects communication\textemdash is 
a well-documented phenomenon in spoken and written language, its presence, consequence, and significance has thus far been little studied in data visualization. Although a nascent body of work has productively begun to broaden the field’s definition of effectiveness beyond affording fast and accurate decoding of propositional meaning to account for contextual factors (including framing, affect, and personal background or disposition), little attention has been paid to how visualizations may communicate an entirely different \emph{type} of meaning that goes beyond the depicted data values and takeaway messages.

In this paper, we offer empirical evidence that, akin to natural language, visualization has a socio-indexical function: beyond communicating propositional insights about data, visualizations 
convey ideas about 
their social provenance because 
formal features of a visualization (e.g., typography, colour, chart type, and data complexity) can index types of people, situations, and qualities.
Through a series of semi-structured interviews, we show that readers make rich and varied social attributions:
value-laden socio-indexical inferences about the identities and characteristics of the actors\textemdash human and nonhuman\textemdash 
involved in a visualization's production and circulation. For example: 

\begin{displayquote}

``Whoever made this is not a researcher, not a scientist. They’re not necessarily going to know the difference between a study published by Dr. Oz and a study published by like Yale Medical.'' 

``Naïve. I feel like the person who made this has the least idea data could be presented differently to convey the story they want to tell [...] so it makes it feel like the person has better intentions. 
''
\end{displayquote}

In our interviews, participants discussed both real-world visualizations gathered from a variety of sources 
as well as ``message-obscured'' visualizations (the same images with the text altered to obscure its propositional content). Through this, we demonstrate that readers can make social attributions based not only on content (i.e., data topic or takeaway message) but on design features alone (e.g. chart type, font, etc.). From the design features of a visualization, readers draw indexical inferences about dimensions of identity such as the designer's age, gender, profession, and political orientation, 
as well as design tools used and likely publication venues. Depending on the reader's own identity and social positioning\textemdash which informs their attitudes toward those people, tools, and contexts\textemdash these attributions give rise to further 
inferences about character traits such as personality, competency, intentions, and trustworthiness.
Together, these constellations of social attributions have the power to frame how visualizations are received independent of the specific data values and takeaway message they depict. 

We synthesize these results into a conceptual model (\autoref{fig:results:diagram}) summarizing the relationship between the concepts we introduce as central to the socio-indexical function of visualization.
Specifically, we illustrate how, influenced by a readers’ own sociocultural context (5-B), socio-indexical readings establish chains of associations between: visualization design features (5-A); particular people, groups, tools of production, and modes of distribution (5-C1); and qualities and characteristics further attributed to those people and things (5-C2); that in turn give rise to the reader’s subsequent reception and behaviour (5-D). These chains of association reveal that design features are more than adornments or containers for data. They can signify more than the information designer intended to impart. From typography to chart type, design choices cannot be reduced to mere embellishments\textemdash  superficial aesthetic
additions\textemdash nor semiotically neutral ``best practices'' for accurately representing certain kinds of data. Instead, visualization design features have the potential to be rich with socio-indexical meaning, their presence able to evoke social personae, contexts, and qualities that can affect how people respond to and engage with data visualizations.

\color{black}
\section{Background and Related Work}
\label{sec:background}

Our work joins recent efforts to broaden the scope of visualization research and design beyond the encoding-decoding process. In this section, we situate our contributions with respect to this prior work and describe the conceptual background we draw on from linguistic anthropology, sociolinguistics, and semiotics.

\subsection{Beyond Encoding-Decoding in Visualization Research}
\label{background:vis}

For the last twenty years visualization researchers have called for more nuanced criteria for evaluating the effectiveness of visualization design, beyond affording fast and accurate interpretation. 
Indeed, recent work argues that there can never be general measures of effectiveness, because any evaluation can only be made in the context of a particular communicative objective ~\cite{Fox_VisPsych_theoriesmodels_2023, kennedyEngagingBigData2016}. 
In response, the community has shown growing interest in a broader range of communicative objectives and measures of effectiveness in studies of 
rhetoric ~\cite{markantCanDataVisualizations2022a, hullmanVisualizationRhetoricFraming2011b, kostelnickVisualRhetoricData2007}, 
 emotion~\cite{lanAffectiveVisualizationDesign2024a, lee-robbinsAffectiveLearningObjectives2023a}, storytelling~\cite{segelNarrativeVisualizationTelling2010, kosara2013storytelling}, 
 and trust ~\cite{elhamdadiHowWeMeasure2022a, elhamdadiVistrustMultidimensionalFramework2023a}. 

\label{par:background:vis:engagement}

Exploring this broader range of communicative objectives 
requires that researchers ``zoom out'' from measuring aspects of graphical perception and cognition to investigate a more general class of behaviour: engagement, 
described as ``the processes of looking, reading, interpreting and thinking that take place when people cast their eyes on data visualisations and try to make sense of them'' \cite[pg.2]{kennedyEngagingBigData2016} (see also \cite{mahyarTaxonomyEvaluatingUser2015}). 
Engagement research has explored how narrative \cite{hullmanVisualizationRhetoricFraming2011b} and persuasive \cite{markantCanDataVisualizations2022a} functions of visualization 
interact with conventions that mandate clarity and minimalism \cite{kostelnickVisualRhetoricData2007},
and how use of embellishment influences outcomes like trust \cite{alebriEmbellishmentsRevisitedPerceptions2024}. 
Similarly, analyses of visualization commentary has revealed readers of data-centric
blogs \cite{hullmanContentContextCritique2015} and subreddits \cite{kauerPublicLifeData2021a} 
assess visualization creators' design decisions and 
intentions, as well as the structure and data-content of the artifacts under critique.  In addition, recent qualitative research has expanded our understanding of how non-expert audiences receive visualizations. 
Interviews with rural Pennsylvanians demonstrated  
engagement is deeply grounded in an individual's assessment of an artifact's personal relevance 
\cite{peckDataPersonalAttitudes2019a}, and interviews with Canadians active in public spaces 
revealed identity-based differences in openness to interaction
\cite{heEnthusiasticGroundedAvoidant2024}.  
He \& colleagues introduced the term \textit{information receptivity} to describe a viewer's inclination to engage with visualizations about a particular topic, in a specific 
context, at a given time~\cite{heEnthusiasticGroundedAvoidant2024}. 
This research on engagement offers converging evidence that understanding how and why people interact with visualizations necessitates an examination of their sociocultural contexts. We join this chorus by offering a way to ``zoom in'' one level of abstraction: connecting viewers' responses to a visualization to their socioculturally grounded associations with particular design features which\textemdash if consistent with the socio-indexical function in 
language\textemdash we expect to influence engagement and reception. 

\label{par:background:vis:aesthetics}

Our attention to the relationship between design features, engagement, and reception also complements the community's renewed interest in visualization aesthetics. Early guidelines for visualization often construed aesthetics as extraneous to an artifact's function as a vehicle for insight; with design decisions other than the most perceptually-discriminable forms of encoding characterized as \textit{chartjunk}
\cite{Tufte1985}. Modern perspectives acknowledge
a more dynamic interplay between form and function, 
with conceptual models differentiating between affective and reflective responses to aesthetics \cite{Leder2004}, showing how aesthetics can realize intended purposes 
\cite{lauModelInformationAesthetics2007}, 
challenging the reductive dichotomy between functionality and artistic expression\cite{moereRoleDesignInformation2011} 
Research has demonstrated that 
`discretionary' embellishments can in fact positively affect how an artifact is received ~\cite{harrisonInfographicAestheticsDesigning2015, alebriEmbellishmentsRevisitedPerceptions2024}, enhancing recall without impairing comprehension ~\cite{batemanUsefulJunkEffects2010, borkinMemorabilityVisualizationRecognition2016}, 
and leading readers to judge a visualization as more enticing ~\cite{alebriEmbellishmentsRevisitedPerceptions2024}. 
Joining the field's move away from terms like chartjunk toward a more capacious understanding of the form-function relationship, 
we attend to how research participants make value-laden assessments of a  visualization based on the people, tools, and contexts with which they associate specific design features.  
Furthermore, we use the term `design features' to include all representational choices made in the development of a visualization (including chart types and encodings), rather than solely `discretionary' embellishments.

\subsection{The Semiotics of Communication as Social Practice}
\label{sub:background:lingath}

Modern visualization research has incorporated semiotics\textemdash the study of signs and meaning-making\textemdash throughout its history, and we join 
colleagues who leverage semiotic theory for inquiry into human factors of data communication. \cite{Bertin1983, Ware2004, hullmanVisualizationRhetoricFraming2011b, FoxHollan_VisPsych_researchprogramme2023, aielloInventorizingSituatingTransforming2020}. We argue that semiotician C.S. Peirce's triadic distinction between three different ways that a sign can signify\textemdash i.e., make meaning\textemdash offers crucial insight for contemporary problems in visualization research \cite{peirce-hoopes}. Two Peircean signifying functions have been widely studied in visualization (albeit not always using Peirce's terms): \textit{symbolic} signs that signify by agreed-upon convention, such as numerals, text, and coordinate systems; and \textit{iconic} signs that constitute the backbone of modern visualization and signify based on resemblance to their referent, appearing whenever designers include figural elements resembling the data topic and, more foundationally, via mappings of number and association to arrangements of marks in space \cite{tverskyCognitivePrinciplesGraphic1997, Bertin1983}. However, \textit{indexical} signs, which are the focus of our work, signify based on real or imagined causal connection or co-occurrence, and have not been widely addressed in visualization research \cite{schofieldIndexicalityVisualizationExploring2013a}, and the indexical signification of \textit{social} meaning, specifically, has yet to be empirically explored.

By contrast, indexicality and social meaning have been core concerns of modern linguistic anthropology, opening the door to understanding how language use in everyday communicative situations involves both the symbolic signification of propositional information and the \textit{indexical} signification of social contexts, identities, and relationships. From the Enlightenment onward, Western discourse about language and communication has emphasized the symbolic, referential function of the sign at the exclusion of all else. According to this conduit model of communication \cite{reddyConduitMetaphorCase1993}, the core function of language is to transmit propositional information through arbitrary, but conventionalized, signs from the mind of the speaker to that of the listener \cite{reddyConduitMetaphorCase1993, gal2015politics}. However, sociocultural studies of language 
challenge this view as a ``semantico-referential fallacy'' \cite{silverstein1979ideology}, arguing it obscures the importance of social (non-propositional) meaning in understanding pragmatics\textemdash how language is actually used and experienced in practice \cite{blomSocialDeterminatesVerbal1972}. Communicating semantico-referential (propositional) meaning is only one of speech's key functions  \cite{jakobson1960poetics}. Speech can also function \textit{socio-indexically} to signify identity and how speakers are related to each other in the social context of use, an insight that provides ``the fundamental basis of linguistic anthropological analysis and theory today'' \cite{nakassisLinguisticAnthropology20152016}. 
We extend this critique of pervasive semiotic ideologies \cite{keaneSemioticIdeology2018} to suggest that the focus on the accurate and efficient transmission of semantico-referential content in visualization similarly obscures how 
socio-indexical inferences may condition readers' responses to a data visualization. 

As philosopher Mikhail Bakhtin writes, “All words have the "taste" of a profession, a genre, a tendency, a party, a particular work, a particular person, a generation, an age group, the day and hour. Each word tastes of the context and contexts in which it has lived its socially charged life” \cite[pg.8]{bakhtin1981discourse}. The socio-indexical function of language thus refers to how certain registers, dialects, sociolects, accents, genres, jargon, and ``slang'' point back toward their social origins: the contexts in which they are presumed to be used and the characteristics of the people presumed to use them  ~\cite{eckert1999beltenhigh, aghaStereotypesRegistersHonorific1998a, irvine2001style, johnstone2008indexicality}.
In this paper, we explore how design features of visualizations similarly retain the “taste” or\textemdash in the contemporary terms of our participants\textemdash the “vibes” of those contexts, identities, and qualities.

Socio-indexical inferences have profound consequences for communication in practice \cite{hallIdentityInteractionSociocultural2005, mendoza-dentonLanguageIdentity2004, woolardCodeswitching2004}. Research on language attitudes studies this phenomenon 
through matched and verbal-guise techniques \cite{garrett2010attitudes, gilesbillings2004attitude}. Using semantic-differentials, Likert scales, and free responses, participants in these studies offer evaluations of a speakers’ qualities\textemdash such as kindness, honesty, and intelligence\textemdash that differ depending on the speaker's dialect, even if
the words spoken are the same ~\cite{bilaniuk2003gender, ladegaard2000language, lambert1960evaluational, woolard1990changing}. Such inferences about speaker identities can influence how people interpret and respond to speech. For example, studies have demonstrated that audience compliance with requests differ based on the accent in which the request is delivered, and the participants' assumptions about the social groups who commonly use that accent \cite{bourhis1976language}.

More recently, linguistic anthropologists have demonstrated that these semiotic processes are relevant to visual forms of communication, in addition to spoken language. Empirical research has demonstrated that  aspects of visual communication\textemdash including images ~\cite{nakassisLinguisticAnthropologyImages2023}, script \cite{choski2021script}, 
typography ~\cite{murphyFontroversyHowCare2017},  graphic designs ~\cite{murphyFakeNewsWeb2023a}, visual aesthetics \cite{nakassisLinguisticAnthropologyImages2023}, and even chart types\cite{donzelli2024chartsJLA}\textemdash are not semiotically neutral, but can be laden with socio-indexical meaning. Typefaces, for example, link "aesthetics of letterforms to ideas about what kinds of actors use them and what kinds of discourse those typefaces are good at representing" \cite[pg.8]{murphyFakeNewsWeb2023a}. In the present work, we build on this research to introduce data visualizations as a further modality of visual communication with semiotic capacity for engendering socio-indexical meaning.

\section{Exploring Socio-Indexical Meaning in Visualization}
\label{study1-methods}

To address the question of socio-indexicality in visualization, we leverage qualitative methodologies best suited for the exploratory study of sociocultural phenomena: ethnographically-grounded interviews of members of a specific sociocultural group \cite{levyPersonCenteredInterviewingObservation2015}. We conducted (n=15) semi-structured interviews 
over the course of 9 months, where respondents were asked to describe their impressions of a series of visualizations. In the following sections we describe how participants were recruited, how visualization stimuli were selected, and the adaptive interview structure and process of thematic analysis.
\textit{The \href{https://doi.org/10.17605/OSF.IO/ERC6P}{supplemental materials} include the full set of stimuli and interview guide.}

\subsection{Participants} 
\label{study1-methods-context}

Following research on socio-indexicality in verbal and visual language \cite{eckertVariationIndexicalField2008}, 
we expected that any socio-indexical readings of 
visualizations would not be universal, but specific to the background and experiences of particular readers, and the communities to which they belong. Knowledge of a community facilitates more exact, rigorous elicitation  and recognition of themes emerging from socioculturally patterned phenomena; thus, we chose to recruit participants in a 
sociocultural milieu where the first author had conducted extensive  ethnographic research: the social media platform, Tumblr. Following a case-study logic 
appropriate to documenting the existence of sociocultural phenomena, participants were “by design, not representative” 
\cite[pg.24]{small_how_2009} 
of a population at large, but rather selected for their familiarity with a shared set of social, aesthetic, and political norms acquired through extensive engagement with Tumblr. Among social media scholars, the platform's users, and the wider digital public, ``Tumblr'' is not merely a communication platform, but a social identity associated with a unique\textemdash albeit not monolithic\textemdash Tumblr ``culture'' shared by long-term participants \cite{michelle-diss}. Tumblr is widely associated with feminist, queer, trans, and social justice oriented individuals, communities, and conversations 
\cite{dame_making_2016, mccracken_tumblr_2017, renninger_where_2015}. It is characterized by aesthetic, linguistic, and epistemological sensibilities that reject rhetorical and scholastic elitism in favour of equally celebrating both prestigious and non-prestigious styles, tastes, and ways of knowing 
\cite{ashley_tumblr_2019, tiidenberg_tumblr_2021, michelle-diss}. Research has 
documented strong linkages between stylistic preferences and social identity among Tumblr users \cite{michelle-diss}, though we did not know whether this phenomenon would extend to data visualizations, which are relatively uncommon on the platform.  

We recruited participants via an announcement on Tumblr 
yielding 223 responses to an interview recruitment survey 
asking about experience with Tumblr and data visualization. All respondents met our eligibility requirement of 3+ years on the platform,
and ( n=15 ) participants were interviewed based on the order in which they submitted their surveys, scheduling availability, and an effort to represent varying familiarity with data visualizations (from occasional viewers on social media to academics in data-driven disciplines). Incidentally, participants represented a diverse range of demographics in terms of 
age (early 20s to 60s), educational attainment (bachelor’s degree to doctorate), and experience on Tumblr (5-12 years). Participants were compensated \$15 USD/hour, and interviews were 40-90 minutes.

\subsection{Stimuli}
\label{study1-methods-stimuli}

We developed a corpus of 20 visualizations with the goal of optimizing heterogeneity in publication source, 
chart-type, data topic, and our subjective assessment of
aesthetic style and extent of embellishment
\cite{alebriEmbellishmentsRevisitedPerceptions2024}.
We began by sampling the MASSVIS dataset \cite{borkinWhatMakesVisualization2013}, adding images 
from social media to achieve greater diversity in authorship. After the first several interviews revealed that participants could make social inferences from the presented stimuli, we decided to further explore whether design features (in the absence of the context of the underlying data) could also provoke inferences.
For this we developed \textit{message-obscured} 
versions, where titles, captions, and legends were replaced with placeholder text in an equivalent typeface, with attention to matching typographic design elements such as kerning and spatial hierarchy. Although it was not always possible to remove all reference to a visualization's topic given the iconicity of some design elements, alterations did obscure specific arguments and take-away messages. 

\subsection{Interview Structure \& Thematic Analysis}
\label{study1-methods-procedure}

We conducted semi-structured interviews following an ethnographic approach that
intentionally emulates lightly prompted conversation, avoiding a rigid question/answer structure to encourage interviewees to speak freely without worry that responses would be assessed for correctness 
(see \S\ref{study1-results-challenges}) and to allow space for discussion exceeding our predetermined questions. 
We drew on the `evocation phase' of the elicitation interview technique 
\cite{hoganElicitationInterviewTechnique2016} to instantiate an interactional context for answering questions: encountering visualizations on participants' Tumblr ``dashboard''. Using a person-centered approach ~\cite{levyPersonCenteredInterviewingObservation2015} following case-study logic \cite{small_how_2009,yin_case_2002}, each participant 
was presented with different combinations of stimuli and 
slightly different prompts, responsive to  the emergent dynamics of the interview and insights drawn from previous interviews. Interviews were conducted via Zoom 
and began by establishing the aim of the research as exploring people's thoughts about a visualization, rather than evaluating their graph reading ability.
The interviewer then presented a randomly-selected stimulus, and posed a casual, open-ended question such as "okay, so: thoughts about this visualization?"
The interviewer progressively elaborated follow-up questions based on the participant's responses (see \S\ref{study1-results-attributions} for an example) and themes surfaced in prior interviews. When the interviewer judged the participant had nothing further to offer, they proceeded to another stimulus, chosen with attention toward maximizing potential for thematic saturation. Each participant was presented with 8-15 stimuli, alternating between aesthetic styles while ensuring each individual saw both untreated and message-obscured stimuli. Throughout the 9 month interviewing process, the author performed ongoing thematic analysis in the tradition of ethnographic fieldwork and grounded theory \cite{charmazGroundedTheoryEthnography2001}. This involved: taking field notes during each interview, identifying emergent patterns, extracting representative quotations to illustrate themes, and iterative reviewing of field notes and prior interviews to revise, clarify, and elaborate upon themes.

\section{Results}
\label{study1-results}

In the following 
subsections, 
we present key themes emerging from our analysis. Leveraging the documented strengths of vignette and representative discourse for reporting ethnographic data and thematic analysis,
\cite{dourishReadingInterpretingEthnography2014a,obrienStandardsReportingQualitative2014,wutichSampleSizes102024}
we illustrate themes by providing quotations from our interviewees, rather than quantitative accounts of how many participants expressed a given idea, because the presented themes surfaced from intentionally open-ended and variable prompts and not static questions that elicit specific numerable responses from every participant, on each topic and stimulus. This accumulation of thematically aligned quotations demonstrates more broadly how visualizations can function socioindexically, whereby: (1) readers make inferences about the identities and characteristics of entities they presume to be involved in a visualization's production and distribution; (2) readers make such social attributions even when the propositional content of stimuli is obscured, 
revealing that design features in and of themselves carry socio-indexical meaning; 
(3) social attributions include both identifications and characterizations about a range of human and nonhuman visualization actors; (4) socio-indexical meaning is socioculturally grounded; and (5) socio-indexical readings influence reception of visualizations. In \autoref{fig:results:diagram} we offer a conceptual model of the socio-indexical function of visualization that summarizes these concepts and illustrates the relationships between them. We conclude with a discussion of the methodological and theoretical implications of challenges in expressing social attributions. \textit{Participant quotations are indicated as (Interview ID : Stimulus ID), where -UN refers to untreated and -OB to message-obscured stimuli.}

\subsection{Readers Make Social Attributions}
\label{study1-results-attributions}

When given open-ended prompts to discuss their thoughts about a visualization, participants readily offered assessments of the artifact (``It’s pretty'') and the data represented (``these numbers were guestimated''), similar to responses documented in research on public visualization discourse ~\cite{kauerPublicLifeData2021a, hullmanContentContextCritique2015}. However, all 15 also offered a form of response that has yet to be systematically documented in visualization research. Ranging from short descriptions to elaborate narratives, participants articulated nuanced \textbf{social attributions}: socio-indexical inferences about the identities and qualities of the actors they presumed to be involved in a visualization's creation and circulation.

To illustrate our adaptive interview structure 
we give an extended example from our first interviewee, Ann (IN-1), who offered wide-ranging impressions of 15 visualizations. A decade-long user of Tumblr, Ann is a lawyer: highly educated, data-literate, and a critical consumer of information. Her comments reflect this, and she offered sharp assessments of the artifacts' clarity and aesthetic appeal\textemdash or lack thereof. However, in addition to those assessments, the conversation was replete with commentary of a very different sort:

\begin{displayquote}
INTERVIEWER: Alright, here is the next one… [Interviewer pulls up the unobscured version of stimulus \hyperref[fig:s1:results:reddit]{S4} 
then trails off to let Ann fill the silence organically.]

ANN: Huh. I mean, at first, the image, you kind of expect it’s going to be more conservative, but it’s not. [...] It seems it has a very Boomer conservative vibe, but the data is the opposite of that [it portrays the Trump administration negatively] [...] But it’s not professionally done at all [...] It looks like someone put this together in Excel; but he doesn’t really know how to make charts in Excel.

INTERVIEWER: What gives that impression?

ANN: As also being someone who doesn’t know how to make charts in Excel ((laughter)) [...] I’d recognize the slanted lines, its labels, anywhere. [Ann goes on to describe other design elements, including font size and the spacing of the legend that remind her of Excel.] 

INTERVIEWER: So would you share this, or something like this?

ANN: No. [...] I don’t know who would share it. [...] Maybe some stupid person on Reddit? It has kind of like a Reddit vibe to it. But I don’t know–I’m trying to think why.

INTERVIEWER: I would be very interested if you can pinpoint why!

ANN: It has like a faux academic-ness to it. [...]  It’s not being like an infographic and trying to look pretty. [...] It’s trying to be more, not pretentious but [...] the person who made it, you know, thinks they’re cleverer than they are.
\end{displayquote}

\noindent Throughout the interview, Ann 
repeatedly volunteered comments about the "feel" or "vibe" of  visualizations, reading their socio-indexical meaning\textemdash the contexts, personae, and characteristics evoked\textemdash rather than 
data or takeaway messages:

\begin{displayquote}
``This is giving middle or high school. One of those teachers that's not bubbly, but outgoing. Has that PBS Magic School Bus, Scholastic Book Fair kind of vibe. Like 'Look, we're young and hip!', trying to appeal to students [but] knowing you're a little corny'' (ANN : S11-UN)

``This one has more of a scientist feel.'' (ANN : S17-UN)

\end{displayquote}

\noindent 
This emerged as a persistent pattern throughout the subsequent interviews. When asked to speak freely about a visualization, participants consistently made value-laden indexical inferences about a visualization's social provenance. Importantly, while all interviewees provided such social attributions for several visualizations, 13 of 15 did not offer attributions for at least one other visualization (see \S\ref{study1-results-challenges}). This indicates that any priming from the interview structure was not strong enough to generate spurious social attributions. Social attributions only emerged if a visualization did have socio-indexical meaning for that participant.

\subsection{Social Attributions Arise From Design Features}
\label{study1-results-design}

Our results demonstrate design features are integral to how readers construct inferences about the social provenance of a visualization. That is, socio-indexical meaning can be derived from a visualization’s design features above and beyond its propositional content (i.e. the topic of the encoded data, its takeaway messages, etc.) We construe \textit{design features} broadly as any element of a visualization where a representational choice was made, including structural features such as chart type and visual encodings, as well as features more typically described as aesthetic or discretionary (typography, use of whitespace, spatial layout, etc.).

\begin{figure}[h]
 \centering 
  \includegraphics[width=1\columnwidth,
   alt={On the top left of this image is a line chart sourced from The Economist. On the right, is the same visualization but with its text and labels redacted. Below both the untreated and message-obscured versions are quotations of social attributions made by interviewees. Importantly, the social attributions are equally vivid and detailed, regardless of if they were made for the original or obscured visualization.}
   ]{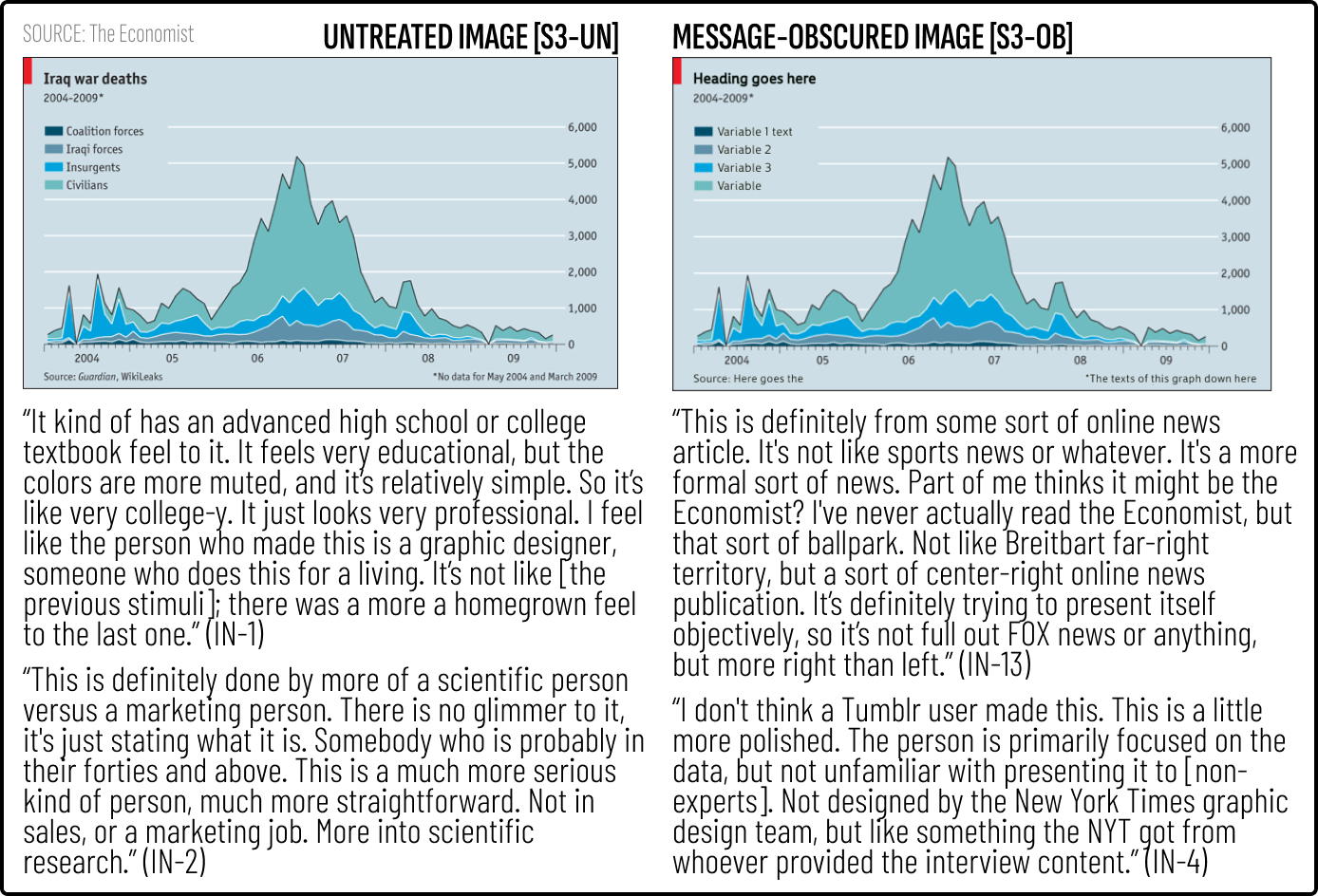}
 \caption{
 \textbf{Attributions of Untreated and Message-Obscured Stimuli}
 }
 \label{fig:s1:results:flags}
\vspace{-2mm}
\end{figure}

We observed that respondents made equally rich social attributions when presented with untreated or message-obscured visualizations, confirming that design features themselves can carry socio-indexical meaning. Consider the similarity in elaboration between responses from interviewees presented with the untreated stimulus in \cref{fig:s1:results:flags} (left) and responses to its message-obscured equivalent in \cref{fig:s1:results:flags} (right).

\vspace{-1mm}

\noindent If readers’ assumptions regarding the entities involved in a visualization's production were entirely determined by its message, they likely could not make social attributions in the absence of that 
propositional content. Moreover, even when presented with the untreated stimulus, IN-1 still credited design features (colour and simplicity)\textemdash not the data topic or message\textemdash with motivating her attributions. In fact, IN-13, presented with the obscured visualization, made a more technically accurate social attribution than participants presented with the untreated stimulus, correctly identifying \textit{The Economist} as the source. Despite never having read the publication, IN-13 links impressions of mainstream formality and an effort to seem "objective," with center-right news outlets exemplified by the \textit{The Economist}. This suggests that socio-indexical readings, far from being separate or even opposed to data literacy, may be a crucial part of it. We conclude that design features alone are capable of conveying socio-indexical meaning. Just as formal features of speech point toward the contexts in which they are presumed to be used and the characteristics of the people presumed to use them, design features of a visualization can similarly index social categories, contexts, and identities. 
However, also like natural language, the socio-indexical meaning of particular design features are not universal. In \S\ref{study1-results-socioculturalcontext}, we discuss visualizations that evoked the most consistent responses among our research participants and how these responses reflect the particular sociocultural context our participants share. 

\subsection{Readers Identify \& Characterize Visualization Actors}
\label{study1-results-attributions:framework}

During thematic analysis we attended to patterns in the attributions participants made, identifying: 
(1) three categories of actors that interviewees commonly implicated in the production and distribution of a visualization (\textit{makers}, \textit{tools}, and \textit{modes}) and  (2) two distinct forms of social attribution (\textit{identifications} and \textit{characterizations}). 
\subsubsection{\textit{Social Attributions Encompass Makers, Tools, and Modes}}
\label{study1-results-attributions:actors}

\begin{figure}[h]
\vspace{-1mm}
  \centering 
   \includegraphics[width=1\columnwidth,
   alt={This table lists the common categories of actor that were referenced by research participants: maker (subdivided into principal, author, and animator), mode, and tool. Next to each category is an example of a social attribution made for that category of actor, such as "done in Google sheets" for Tool and "some poor intern" for author.}
   ]{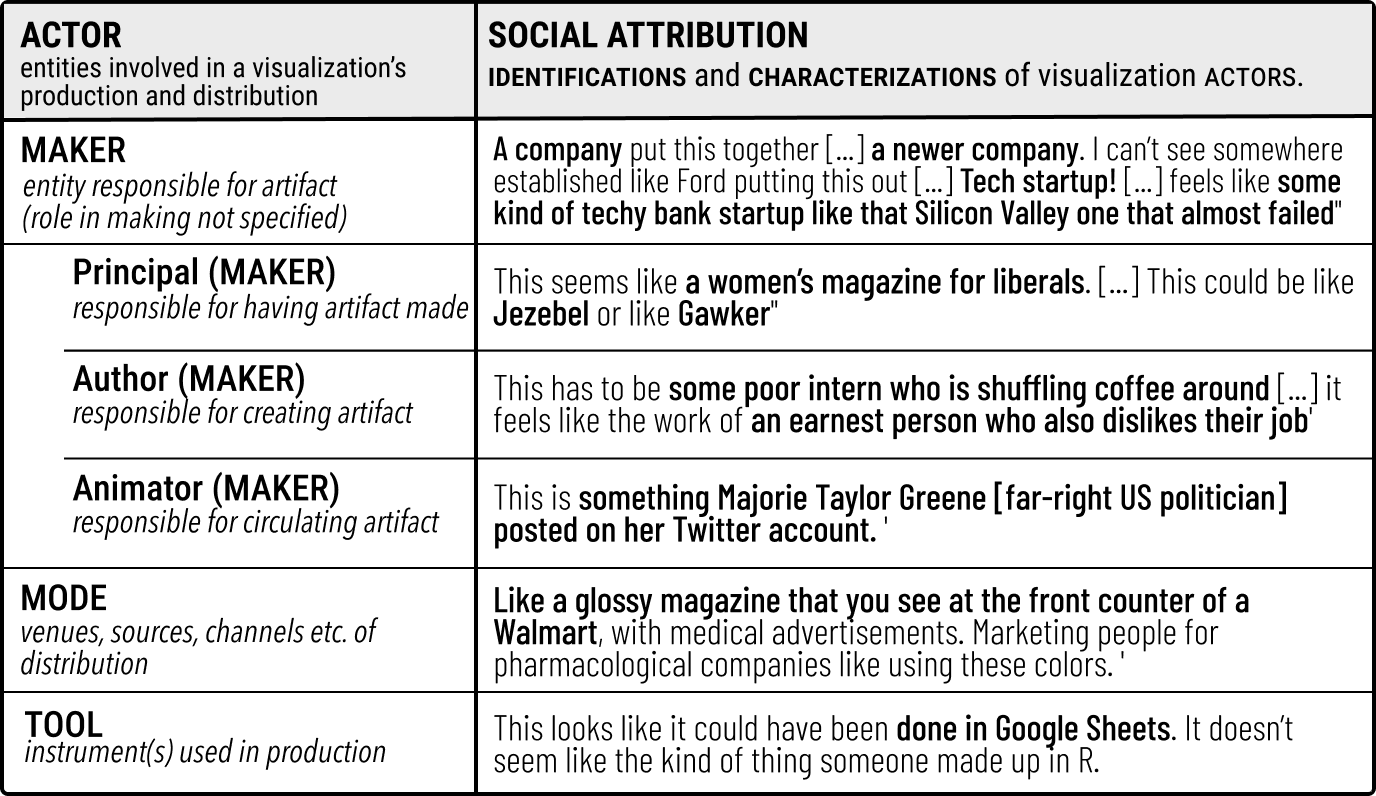}
  \caption{
  \textbf{Examples of Social Attributions of Visualization Actors.} 
  }
  \label{fig:s1:results:examples}
  \vspace{-4mm}
\end{figure}

Consistent with the results of sociolinguistic guise studies showing that listeners evaluate speaker qualities based on the formal features of their speech ~\cite{bilaniuk2003gender, carrie2018american, ladegaard2000language}, we found that interviewees readily drew inferences about the imagined creator of a visualization, which we categorized as \textbf{maker}, referring to properties such as age, gender, education, occupation, and political orientation. However, in contrast to sociolinguistic studies where inferences focus on a single speaker, viewers made social attributions about several kinds of entities that could be involved in an artifact’s production and circulation, including collective entities as well as
\textbf{tools} for creation and \textbf{modes} of distribution. We categorized these entities\textemdash individuals, institutions, means and methods of production and distribution\textemdash under the umbrella of \textit{actors}, to underscore the importance of recognizing both human and nonhuman agents in the emergence of material and semiotic artifacts 
\cite{latour1987action, sayes2014actor}. 
Review of interview field notes alongside transcripts further revealed that social attributions about 
the entities we originally categorized as makers
could be further subdivided 
using categories developed by sociologist Erving Goffman \cite{goffmanFormsTalk1981a}
to reflect actors involved in different aspects of the production process in communication: \textit{principal}, the person or collective responsible for the commission of the artifact; \textit{author}, the person responsible for materially creating the artifact; and \textit{animator}, the person responsible for distributing or sharing the artifact. \autoref{fig:s1:results:examples} defines each category of actor we identified, alongside example social attributions from our participants.

\begin{figure*}[t]
 \centering 
  \includegraphics[width=1\linewidth,
  alt={Figure 4 has two rows. The first is dedicated to stimulus S4-OB, a line chart with a gradient that goes from yellow to red to black and a faded American flag in the background. To the left of it are five quotations from different research participants where they all identify the social media platform, Reddit as an actor the visualization evokes. To the left of that, there are five additional quotes which describe the characteristics these participants associate with Reddit and its users. The row below mirrors this structure. An image of S1-Ob characterized by a "millennial pink" background and bar chart made of foliage sits besides participant quotes identifying it with a young, female, social media content creator.}
  ]{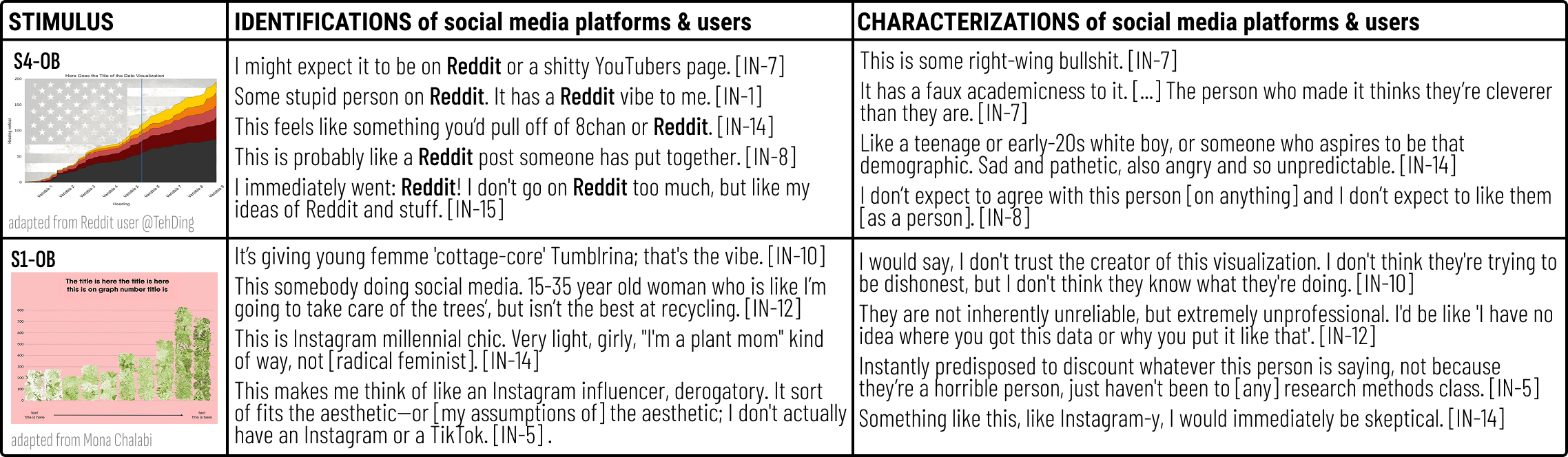}
 \caption{
 \textbf{
 Sociocultural groups draw on shared indexical fields, leading to similar identifications and characterizations}
}

 \label{fig:s1:results:reddit}
\vspace{-6mm}
\end{figure*}

\subsubsection{\textit{Attributions Include Identifications and Characterizations}}
\label{study1-results-attributions:identifications}

In addition to categorizing common types of actors interviewees implicated in the production of a visualization, we also noted two distinct forms of attribution. The first involved \textit{identifications} of particular people, groups, or social categories, including remarks about properties such as age and gender (``Someone sort of mid-20s, I’d guess female”), occupation (``Someone’s admin assistant”), or specific source (``The Washington Post”). The second entailed \textit{characterizations} of traits, qualities, or values, including statements about personal disposition (``angry and unpredictable”), assessments of competency (``Don’t know what they’re doing”), and a range of explicitly evaluative adjectives (``biased”, ``genuine”, ``trustworthy”).

Importantly, identifications and characterizations can be equally laden with social meaning\textemdash entangled with and reflective of social norms, values, categories, contexts, experiences, and relationships. Maker identifications may appear, on the surface, more value neutral than characterizations, which make direct evaluative claims about a maker (e.g. “passionate” or “lazy”). To identify a maker as “in their 30s”, “white lady”, “an intern”, or “FOX News” does not explicitly express a social evaluation of that actor. However, identifications can imply a host of associated characterizations. For example, ostensibly, generation identifications simply mark when a person happened to be born. However, from “OK Boomer” memes to articles critiquing "Gen Z work ethic", these labels often index far-from-neutral qualities and characteristics (e.g regressive social attitudes or laziness, respectively). Seemingly neutral tool and venue identifications can be similarly value laden. IN-13 describes stimulus \hyperref[fig:s1:results:reddit]{S1-OB} as “a very sort of Instagram type of graph”, which could be a straightforward naming of the venue in which the visualization might have been published. However, when our interviewer asked what they meant by that description, IN-13 went on to describe “Instagram type” with a series of characterizations they associate with the platform, its users, and its content: less “masculine”, focused on “style over substance”, and “not aimed at computer nerds” (see \S\ref{study1-results-socioculturalcontext} for further discussion). Similarly, when IN-4 said S12-OB “has Microsoft Word vibes”, she elaborated “but not necessarily like in a bad way”, implying that while being created with the tool Microsoft Word can index qualities with negative connotations (e.g. a lack of sophistication and beauty), in this case those qualities are positively linked with sincerity or authenticity: 

\begin{displayquote}
“If it’s so pretty that it’s clearly been made by a graphic designer and approved by a marketing team, it’s trying to sell you something. But this? Microsoft Office vibes are very neutral and a bit more objective.” 
\end{displayquote}

\noindent
It is therefore crucial to recognize that seemingly neutral identifications can function as conceptual shorthand for constellations of value-laden characteristics. Identities and characteristics are linked to each other\textemdash and to additional identities and characteristics\textemdash through chains of socio-indexical meaning. The socio-indexical significance of a visualization is rarely a direct one-to-one relationship between a single feature and a single attribution, but rather comes into being through elaborate entanglements of aesthetics, attitudes, and associations within a broader, culturally-specific field of indexical meaning \cite{eckertVariationIndexicalField2008}.

\subsection{Sociocultural Context Influences Attributions}
\label{study1-results-socioculturalcontext}
\color{black}

Our results illustrate that social attributions are contextual and culturally patterned, resulting in inferences that vary between individuals, but also reflect the broader cultural attitudes and beliefs of the communities in which readers participate. 
In general, our results showed significant variance in the social attributions made for most stimuli, but
S1 and S4 garnered remarkably consistent and specific responses among the 8 interviewees with the most familiarity and alignment with Tumblr. The bottom row of \autoref{fig:s1:results:reddit} documents how, for these interviewees, S1's visual simplicity, figural leaf embellishments, `millennial pink' and overall muted pastel colour palette consistently indexed a Gen Z or Millennial woman who creates content online in places the interviewees associate with `style over substance' (e.g. Instagram, Pinterest, the ``cottage-core'' corners of Tumblr), and whose politics could be described as more liberal than conservative, but who does not engage in enough critical thinking to have developed more actively anti-racist and anti-capitalist stances and practices. In contrast, (top row) S4's flag imagery,  ``aggressive'' saturated colours, and ``unpolished look'' consistently indexed the social media platform Reddit and interviewee's perception of its stereotypical user: an angry young man who couches his far-right views in the rhetorical style of scientific authority.

This cohesion reflects the shared experiences and ideologies of those Tumblr users who strongly self-identify with being part of a ``tumblr'' community, and thus with the feminist, anti-racist, LGBTQ+, and leftist perspectives, and various forms of aesthetic and rhetorical expression, that flourish on the platform. A central part of identity formation (individual and group) involves defining ourselves in contrast to others around us\textemdash or at least our perceptions of others around us. Be it countries, religions, or fandoms, ``we'' are largely defined as ``not them'', which requires constructing robust imaginaries of ``them'' \cite{becker1963labeling}. For dedicated Tumblr users, the ``others'' against which they understand themselves are often different types of social media users. Therefore, for individuals in this sociocultural milieu, 
``Instagram'' and ``Reddit'' are easily made and, most importantly, indexically rich identifications that further imply a host of heavily value-laden characteristics that influence interviewees' reception of the visualization\textemdash as illustrated in the last column of \cref{fig:s1:results:reddit}. In contrast, non-Tumblr survey respondents in a follow-up study we conducted similarly described the design features of these stimuli, but did not specifically associate them with Instagram or Reddit (see Fox \& Morgenstern et al. \cite{PAPER2}).

This convergence, reflecting the shared experiences and ideologies of these Tumblr users, substantiates the idea that socio-indexical meanings are not universal, but culturally specific to particular communities of use contexts. IN-11 herself acknowledged that her embeddedness in Tumblr’s cultural context influenced her attributions. At the end of the discussion, our interviewer confirmed to IN-11 that the visualization had been sourced from Reddit, as she had surmised. IN-11 grinned widely and said, ``I don’t know if I should be proud of myself? Or...,'' she trailed off with a self-deprecating laugh before continuing, ``questioning how much time I spend on the internet''. Our interviewees responses to S1 and S4 underscore that socio-indexical readings are always grounded in prior sociocultural experiences, knowledge, or narratives, such that design features can prompt adversarial responses to a visualization if they index identities that a respondent opposes on the grounds of their own social positioning.

\subsection{Social Attributions Impact Trust \& Reception}
\label{study1-results-engagement}

This study relied on self-reported attitudes and did not directly measure engagement behaviour. Nevertheless, 12 of 15 interviewees made remarks implying that, were they to encounter a given visualization in the wild, their socio-indexical readings would influence how they might engage\textemdash or not\textemdash with that visualization.

\begin{displayquote}
``If this were to come across my dash, for, like a [topic] I don’t know anything about, I would automatically distrust that person's opinion on that [topic], I would be like, I don't know [what this is about], but you're probably wrong.'' (IN-4: S14-OB)

``I trust this the most [of all the presented visualizations]. I feel like this comes from a newspaper which would have a political bias that probably sits closer to my own personal political bias.'' (IN-13: S7-OB)

\end{displayquote}

\noindent 
Moreover, this study surfaced potential complications of seemingly straightforward relationships between trust and qualities of a visualization such as beauty \cite{linVisualizationAestheticsBias2021a} and bias. Responses indicated that qualities that might be presumed to carry connotations that are widely regarded as negative or positive, such as bias and beauty, respectively, often had complex relationships with trust. For our participants, there were instances where: beautiful design indexed corporate lies (S5); ``terrible'' design and incompetency indexed authenticity and good intentions (S1, S20); makers viewed as biased was nevertheless assessed as trustworthy because they seem ``closer to my own personal political bias'' (S2); and a maker read as unbiased was nevertheless deemed untrustworthy because they were presumed to also be incompetent. We further address how socio-indexical meaning complicates the relationship between trust and beauty in a series follow-up studies \cite{PAPER2}.

\subsection{Interpreting Challenges in Elicitation}
\label{study1-results-challenges}

Our interviewees made a wide range of social attributions to presented visualizations, but eliciting those attributions was not without challenges. The instances where participants were unable or reluctant to express social attributions (see \S\ref{study1-results-attributions})
highlight three factors that can inhibit the expression of socio-indexical inferences in the context of data visualization: (1) \textit{visualization ideologies} (i.e. beliefs about what data visualizations are, how they communicate, and how readers should interact with them) that make expressing socio-indexical readings a socially precarious behaviour; (2) insufficient\textit{ metavisual awareness} (i.e. the visual equivalent of metalinguistic and metapragmatic awareness\textemdash the ability to consciously identify and describe elements of language and how they are used); and (3) the \textit{indexical salience} of particular design elements for particular individuals.

\subsubsection{\textit{Visualization Ideologies}}
\label{study1-results-challenges:ideologies}

People are enculturated beings, socialized into particular norms of communication \cite{schieffelin_language_1986}\textemdash including norms of politeness \cite{haugh_revisiting_2004} and widespread ideologies about data visualizations that inform the range of socially acceptable responses to probing questions. During interviews, participants made remarks suggesting that at times they felt it inappropriate to make socio-indexical inferences (even while acknowledging that they can and do):

\begin{displayquote}
    “I’d guess female. But that’s terribly sexist of me.”

“Oh, I don’t know how to say this without sounding mean.”

“If [the maker] ever finds that I’ve said this, I’m going to feel so bad.”

“I don’t want to like judge people right off the bat”
\end{displayquote}

\noindent 
Beyond the social undesirability of being prejudiced or “mean”, students in the United States are often taught from a young age that scientific data are objective, not socially-mediated, and practices of formal instruction in STEM education reinforce such ideologies \cite{massoudObjectificationInscriptionKnowledge2008}. 
From this perspective, a maker's identity, the tools used to produce a visualization, and the venue where it circulates should have no impact on how a visualization is designed or interpreted. This makes explicit expressions of the fundamentally subjective socio-indexical meanings explored in this study epistemologically transgressive; the very possibility of socioindexically reading visualizations violates widespread beliefs\textemdash among scholars, practitioners, and audiences\textemdash that data visualizations should be neutral conduits for objective information \cite{massoudObjectificationInscriptionKnowledge2008}.

For example, after characterizing pie charts as ``reasonably trustworthy'' and having ``more neutral vibes'', IN-14 sheepishly acknowledged ``that could be my experience making me a little biased''. 
She went on to cite her workplace as the source of the association, explaining that they often used pie charts
to report straightforward demographic information. This was a wonderfully articulated chain of socio-indexical reasoning\textemdash exactly what this study sought to explore. Yet IN-4 offered it apologetically, as if she had done something wrong. Such discomfort with acknowledging how personal experience influenced assessment of a visualization emphasizes how non-propositional readings are not socially preferred
within culturally dominant frameworks that situate data visualizations as scientifically objective, making such readings more challenging to elicit.

We discovered that the social force of such visualization ideologies could be partially mitigated by 
framing the interview as a casual, friendly discussion where participants would not be evaluated on their performance. Our interviewer did this by: (1) 
directly stating that the point of the exercise was not to determine if participants could “accurately” interpret a visualization; (2) following a semi-structured, conversation-based interview structure; (3) using hedging language in probes (e.g., “In your opinion”); and (4) maintaining a cheerful and affirming demeanour. In some instances, however, participants were willing but still did not make social attributions for a particular visualization, indicating  additional factors that can impact the elicitation of socio-indexical readings: a reader's metavisual awareness and the indexical salience of particular visualizations.

\begin{figure*}[t]
  \centering 
    \includegraphics[width=1\textwidth,
    alt={Figure 5 is a flow diagram illustrating our conceptual model of the socio-indexical function in visualization. It describes that visualization features (box A) consisting of design features and data features, combine with a viewer’s sociocultural context (box B) to give rise to social attributions (box C).  Social attributions are socio-indexical inferences about the actors involved in a visualization's provenance, and come in the form of identifications (box C1) and characterizations (box C2). Finally, social attributions affect reception and behaviour (box D).  Below the diagram there is an example quotation that illustrates each component of the model.  The quotation reads: “This gives me like a Newsweek type of vibe. This doesn't look like a default colour scheme I'm familiar with, so some attention was paid to [design]; it's not as slick as I’d expect in, say, the New York Times, but it's professional. […] This  looks like mainstream media […] so I would be inclined to look into the actual numbers behind it. Because of my personal politics, I am a little dubious of how [some] topics are covered [by this kind of publication].”
    }
    ]{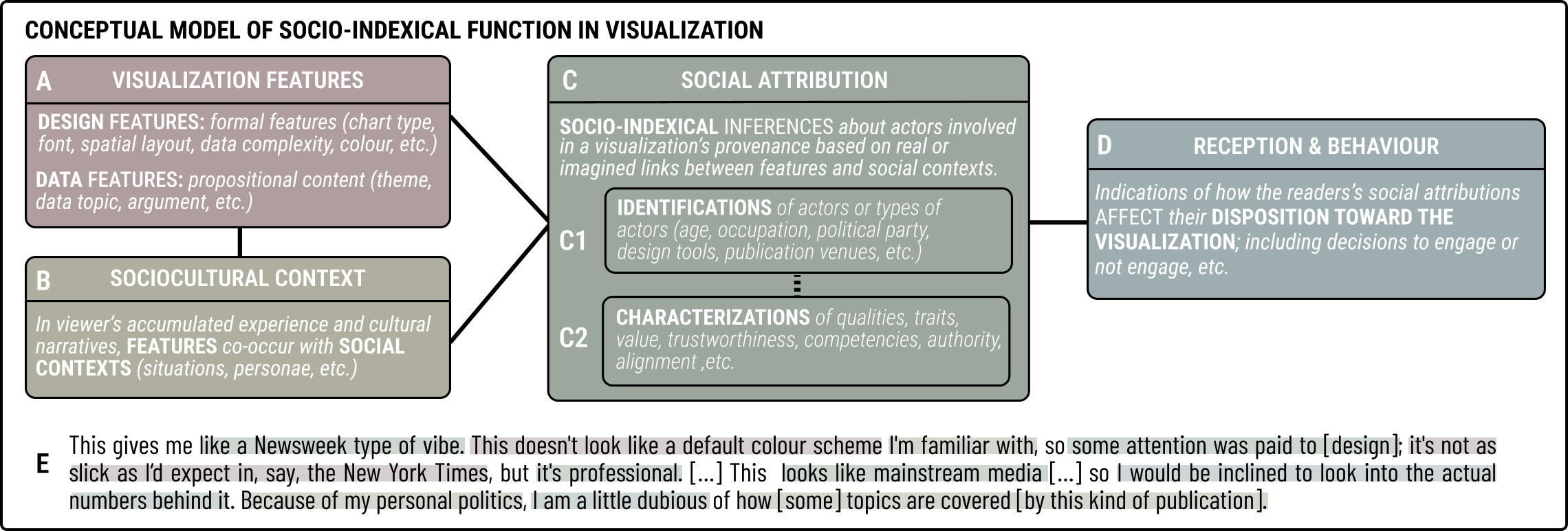}
  \caption{
  \textbf{Conceptual Model of Socio-Indexical Function in Visualization.} 
  The combination of features of the artifact and the sociocultural context of the viewer combine to give rise to social attributions  (expressions of socio-indexical inferences), that in turn impact reception and behaviour. 
  }
  \label{fig:results:diagram}
  \vspace{-4.5mm}
\end{figure*}

\subsubsection{\textit{Metavisual Awareness}}
\label{study1-results-attributions:metavisual}

What do readers attend to when making social attributions? Our results  demonstrate that design decisions can carry socio-indexical meaning (\cref{study1-results-design}), but these findings also suggest that exactly \textit{which} design features readers attend to may often be beyond readers' conscious awareness, as evidenced by how interviewees articulated associations between design features and attributions with varying degrees of explicitness. Sometimes participants did precisely link specific design elements (e.g., colour, typography, chart type, and relative complexity or detail of the graph)  to specific social attributions. For example:

 \begin{displayquote}
    ``Probably someone in their thirties or forties. That font is giving exceedingly strong 80s/90s vibes.'' (IN-5: S2-OB)

   ``It's more scientific because the amount of data […] So I'm less likely to think it's from the IRS, and more likely to think it's [a scientific article].'' (IN-4: S9-OB)

\end{displayquote}
 
\noindent 
More often, participants made attributions without naming the specific design features and just referenced an overall aesthetic (e.g. ``The presentation [of the visualization] makes me go, ‘you [the maker] seem establishment-y’.'' [IN-7: S7-OB]), and at times even expressed uncertainty about what exactly provoked the social attribution (e.g. ``I don't know why I say that?'' [IN-6: S5:OB]).

Metalinguistic and metapragmatic awareness of how ways of speaking are associated with types of people and contexts 
is widespread, with metalinguistic discourse a ubiquitous feature of everyday conversation. However, the kind of \textit{metavisual awareness} we are studying, particularly as it pertains to data visualization, is far more specialized. Simply put, the vast majority of people speak\textemdash and speak about ways of speaking\textemdash more frequently than they create, encounter, and reflect upon data visualizations. Thus, even when interviewees made socio-indexical readings, their capacity to articulate the chains of inference that motivated those readings was often limited. This is exemplified by IN-15 when she precisely links one design aspect of S6-OB to her social attribution, but cannot articulate the intervening inferences: ``Definitely reminds me of a women's health presentation. Huh. [Laughter.] That's weirdly specific! I don't know why this purple and blue is giving women's health.''

\subsubsection{\textit{Indexical Salience}}
Because some visualizations prompted robust social attributions from participants who saw them (e.g., S1 and S4; for 15 of 15 participants), while others did not (e.g., S10 and S12; for only 2 of 5 participants),
we suggest that the likelihood of socio-indexical readings depends on a quality we call \textit{indexical salience}. Just as not every formal feature of speech is strongly associated with a particular social context or identity, not all design choices are strongly associated with particular social attributes\textemdash some are more indexically salient than others. However, this quality is not universal; the design choices that are indexically salient for one person or sociocultural group, might not be for another. It is important to note that 
embellishment should not be assumed to inherently correlate with indexical salience. Interviewees made similarly detailed or vivid attributions for both more 
(e.g. S2, S4) and less (e.g. S12, S16) embellished visualizations. Simplicity itself can index certain makers, tools, and channels (e.g. ``someone trying to look objective'', ``a scientist'', or ``Microsoft Word vibes''). In contrast, a more embellished visualization like S10 was less strongly associated with particular actors (e.g. merely described as “pretty”).

\subsection{A Conceptual Model of the Socio-Indexical Function}
\label{study1-results-model}

Taken together, our findings demonstrate that visualizations communicate more than the data they encode. In addition to a propositional function, visualizations also have a socio-indexical function\textemdash conveying impressions of their social provenance. Our evidence shows how viewers respond to and engage with visualizations in the context of what they imagine designers’ identities to be. We illustrate our conceptual model of this socio-indexical function in
\autoref{fig:results:diagram}, 
describing how the features of a visualization (A) act in concert with a viewer’s sociocultural context (B) to give rise to social attributions (C) with the power to influence reception and behaviour (D). In (E) we illustrate how 
these concepts can be located in interviewee discourse. We offer our model of the socio-indexical function of visualization not as a replacement for information-processing models, 
but an extension that reveals an additional dimension of meaning that can be made with visualizations. As a conceptual model\cite{friggModelsScience2006,jabareenBuildingConceptualFramework2009}, it serves as a tool for thinking about the consequences of design decisions; suggesting new research questions (\S\ref{subsubsec:discussion:implications:research}), 
and interventions in the design process (\S\ref{subsubsec:discussion:implications:design}).

\section{Discussion and Future Work}
\label{sec:discussion}

 Inspired by linguistic anthropological research on socio-indexicality demonstrating that listeners make inferences about speakers from formal features of language\textemdash inferences that impact how listeners engage with an utterance\textemdash we asked whether readers might make similarly non-propositional readings about data visualizations. Our results document the presence and consequence of a socio-indexical function of data visualization, surfacing five key thematic findings: (1) readers can make social inferences about 
 the identities and characteristics of 
 visualization actors (e.g. makers, tools, modes of distribution) (\S\ref{study1-results-attributions}); (2) these value-laden social attributions, positive and negative, can be made even when the propositional content of stimuli is obscured, demonstrating that design features in and of themselves can function as socio-indexical signs (\S\ref{study1-results-design}); (3) socio-indexical meaning is contextual and socioculturally grounded, emerging from chains of associations rooted in readers' own social positioning, experiences, values, and cultural context (\S\ref{study1-results-socioculturalcontext}); and, ultimately, (4) viewers' socio-indexical readings of a visualization can influence reception and engagement (\S\ref{study1-results-engagement}). In \autoref{fig:results:diagram} we offer a diagrammatic representation of this conceptual model, illustrating how readers make inferences about a visualization's social provenance based on their socioculturally grounded attitudes toward the people, tools, and contexts with which specific design features are associated, which in turn influence behaviour. While it remains of utmost importance that readers make accurate (propositional) inferences about visualized data, these results emphasize the crucial role of non-propositional, \textit{socio-indexical} readings in visualization reception. Acknowledging this socio-indexical function offers both theoretical and practical implications for visualization research and design, from prompting new research questions not evident given an information-processing model of communication (\S\ref{subsubsec:discussion:implications:research}), to encouraging designer-driven exploration of socio-indexical meaning for target audiences during the design process (\S\ref{subsubsec:discussion:implications:design}).

 \subsection{Limitations: Measuring Socio-Indexicality}
\label{subsec:discussion:limitations}
In this paper we provide evidence for the existence of a socio-indexical function of visualization. 
While this demonstrates that the phenomenon can occur, it does not enumerate all its possible expressions nor predict the conditions under which it will occur. 
Ethnographically-informed, person-centered interviews are highly effective for in-depth phenomenological study, but like any research method they have limitations. First, our participants’ socio-indexical readings inevitably reflect perspectives rooted in their specific sociocultural context. For example, Tumblr users are known for their countercultural left-leaning politics 
and this influenced their social attributions (e.g. negative characteristics associated with identifications like “The Economist”). 
The structure of research in disciplines like anthropology tells us that sociocultural phenomena\textemdash like socio-indexicality\textemdash can only be studied in the context of specific sociocultural groups because there is no such thing as a ‘general’ population: every individual belongs to intersections of sociocultural groups. As Dourish writes, ``ethnographic work often generalizes, but it does so through juxtaposition contradistinction, comparison, sequentiality, referentiality, resonance, and other ways of patterning across multiple observations.'' \cite[pg.13]{dourishReadingInterpretingEthnography2014a} Thus, demonstrating the broader contours of such phenomena requires an accumulation of studies 
with different populations. We hypothesize such studies would also yield socio-indexical readings, but with variance in the salience of particular design features and specific indexical associations. Second, interviewer efforts to mitigate the social forces constraining elicitation (see \S\ref{study1-results-challenges})
could lead participants to offer social attributions when they might otherwise not, raising questions about the feasibility of eliciting socio-indexical readings asynchronously. We address this in a follow-up study exploring the structure of variance in social attributions via quantitative analysis of attitude-elicitation surveys \cite{PAPER2}. Third, our results reflect only what interviewees are able or willing to verbally articulate (see \S\ref{study1-results-attributions:metavisual}). We therefore recommend implicit measures and situated experimental studies be used to address how socio-indexical readings may affect graphical perception, comprehension, and behaviour in ways that participants cannot consciously express. 
Finally, although this study used both natural and message-obscured stimuli, we did not systematically attend to interplays between data and design. An important next step that we address in subsequent work \cite{PAPER2}, is to explore how indexical and propositional meaning might interact when eliciting inferences with untreated naturalistic stimuli.

\subsection{Implications of the Socio-Indexical Function}
\label{subsec:discussion:implications}

We argue that socio-indexicality has important implications for visualization research and design, as readers’ reception and engagement with a visualization is influenced by not only content or even aesthetic appeal, but also their sense of alignment or disalignment with the actors they imagine to be involved in its production and circulation. This exploratory study thus serves as a starting point and call to action for a much broader programme of sociocultural inquiry in visualization; one with practical implications for both research and design.

\subsubsection{Implications for Researchers}
\label{subsubsec:discussion:implications:research}

At the highest level, our model of the socio-indexical function reveals new research questions that more canonical models of (propositional) information processing do not afford. For example, what other forms of social meaning might design features index?  
\autoref{fig:results:diagram}-C showcases the actor attributions revealed  in this study; however, 
linguistic features can index not only the identities and characteristics of participants in an interaction, but also the type of interaction itself\textemdash 
what sort of situation it is, and therefore what roles and behaviours participants are expected to take up\cite{durantigoodwin1992context}. Just as certain features of natural language can frame an interaction as cooperative, competitive, educational, or playful\textemdash 
encouraging certain behaviours while limiting others\textemdash  might certain visualization design features similarly define a reader’s encounter with an artifact in ways that facilitate certain forms of engagement? 
Part B of the model provokes us to ask, what are the processes (e.g., branding, historical accumulation) through which design elements come to be enregistered as exemplars of particular social characters and characteristics? That is, what experiences and narratives establish associations between design features and social personae in the first place\textemdash and is there space for intervention in the generation of these associations? Each part of the model, and each point of connection between them, offers opportunities for further inquiry.

\subsubsection{Implications for Designers}
\label{subsubsec:discussion:implications:design}

Because the contextual nature of socio-indexicality precludes the generation of universal mappings between design features and meaning that generalize across all contexts and audiences, the actionable design recommendation we offer is a \textit{procedural} intervention, rather than one of specific design guidelines (e.g. the best font to index trustworthiness.) 
In the design process, we suggest designers systematically document associations between design features and social attributions among their target audiences. Visualization designers already account for aspects of audience via designerly ways of knowing \cite{rettbergWaysKnowingData2020}; however, as evidenced by empirically-driven guidelines \cite{franconeriScienceVisualData2021}, greater emphasis is currently placed on the reader’s cognitive-perceptual characteristics than the fields of socio-indexical meaning with which they interact. While such guidelines remain important for affording efficient reading of data, socio-indexical readings can influence if and how an audience engages with a visualization in the first place. Our conceptual model provides a reference for what needs to be discovered to predict the socio-indexical consequences of design decisions, or troubleshoot breakdowns in communication. For example, to communicate scientific health data (D. Behaviour: engage with science communication) to a particular community (B. Sociocultural Context: that community) wary of medical professionals (C1: Identifications: hospitals, doctors; C2: Characterizations: uncaring, arrogant), designers could employ user research to work backward through the model to determine what design features (A. Visualization Features: to be discovered) the community associates with those identifications and characterizations and involve those discoveries in design decisions. For an example from the current study, designers seeking to address long-time Tumblr users might avoid the combination of flags and an “unpolished look”, unless they wish to index “right-wing bullshit” that would be immediately dismissed\textemdash even if the content suggests otherwise (e.g. S4). 
Alternatively, to index sincerity (e.g. S6, S15), they might aim to produce similarly “unpolished” artifacts, but with “softer” colours and minimal iconic elements. 
However, determining precisely what design features this audience sees as contributing to an “unpolished look” (e.g. what fonts, spatial organization, or chart type) necessitates further study. We encourage designers to look beyond perception and cognition when making design decisions, to consider what social meaning their choices convey, and how they might leverage visualization’s robust capacity for engendering socio-indexical meaning, to achieve their design objectives.

\subsection{Sociocultural Visualization Research \& Design}
Our study demonstrates the need for greater attention not only to social and cultural factors in visualization, but to sociocultural \textit{functions} of visualization. 
Our model of the socio-indexical function of visualization echoes Keating \& Javenpaa's ``Communication Plus'' model of language, particularly the core tenet that all language\textemdash including visual language\textemdash is both social and cultural \cite{keatingWordsMatterCommunicating2016}:

\begin{quote}
``In every interaction, language provides its speakers with unique ways to build, rebuild, or destroy relationships between people. […] language use is intertwined with important ideas about what a person is supposed to be (and how a person is supposed to behave as a moral person) and aesthetic aspects of behaviour.'' [pg. 20]
\end{quote}

\noindent The ever-growing sophistication of visual mis- and disinformation and the burgeoning mobilization of data literacy against science itself present challenges for public data communication that cannot be adequately addressed by framing the problem\textemdash and thus possible solutions\textemdash solely in terms of the encoding-decoding process. From such a perspective, the obvious solutions are to improve design and data literacy. Yet, while these are valuable interventions, recent studies of visual misinformation reveal that failing to adhere to best practices in visualization design guidelines is \textit{not} the primary way people ``mislead with visualizations” \cite{lisnicMisleadingVisualTricks2023}, 
and that viewpoints counter to scientific consensus ``do \textit{not} result from a deficiency of data literacy”(emphasis ours)\cite{leeViralVisualizationsHow2021}. Alternatively, a sociocultural approach to visualization emphasizes that effective visualization not only requires attention to how efficiently encoded propositional information can be decoded, but also recognition of the way visual artifacts mediate relationships and identities.
Visualization, like all communication, both conveys information and constitutes social action. A visualization is not just a message, it is also a semiotic \textit{act}, positing and constructing relationships between a range of stakeholders: analysts, designers, publishers, platforms, and ultimately readers. Readers are socially-situated actors, embedded in networks of relationships that shape their personal identities. Of course, whether readers infer visualization actors to hold similar social positions and values to themselves has profound implications for trust. Extending trust to an entity a reader perceives as ``not on [their] side” (IN-2) is a risky venture. Most importantly, however, as receptions of  stimuli \hyperref[fig:s1:results:reddit]{S1 and S4} demonstrate: visualizations can provoke adversarial readings if a reader presumes the maker is not merely someone untrustworthy but, more fundamentally, someone they don’t ``expect to like'', ``don’t expect to agree with”, and\textemdash as these responses suggest\textemdash someone they do not \textit{want} to agree with. 

Depending on what socio-indexical meaning someone reads into a visualization, interacting with that visualization can either reinforce or challenge their sense of socially embedded identity. Particularly 
for divisive issues, public data initiatives that seek to go beyond `preaching to the choir' need to consider that adversarial readings may not be failures of data literacy, but rather enactments of social meaning. For effective public engagement with data, clarity, lack of apparent bias, and appealing aesthetics alone may not be sufficient to reach disaffiliated publics. Indeed, a public hostile to the scientific establishment may well take qualities of visual clarity, objectivity, and beauty as indexical identifiers of groups they oppose. A reader's decisions about how to engage with a visualization certainly concern its data, but perhaps more importantly, deciding if\textemdash and how\textemdash to interact with a visualization can be a statement of a reader's own identities, politics, aesthetics, and dispositions as aligned or disaligned with all the actors they infer to be involved in its production.

\acknowledgments{
  This work was supported by MIT METEOR and PFPFEE fellowships for Amy Fox, an Amar G. Bose Fellowship, Alfred P. Sloan Fellowship and National Science Foundation award \#1900991. 
}

\section*{Supplemental Materials}
\label{sec:supplemental_materials}
Supplemental materials are available on OSF at \url{https://doi.org/10.17605/OSF.IO/ERC6P}.

\bibliographystyle{abbrv-doi-hyperref-narrow}
\bibliography{BIBS/main}

\begin{thebibliography}{10}
\renewcommand*{\sfdefault}{PTSansNarrow-TLF}

\bibitem{aghaStereotypesRegistersHonorific1998a}
A.~Agha.
\newblock Stereotypes and registers of honorific language.
\newblock {\em Language in Society}, 27(2):151--193, Apr. 1998. \href{https://doi.org/10.1017/S0047404500019849}
{doi: \textsf{%
10\hspace{.1pt}\discretionary{.}{%
}{.}\hspace{.4pt}1017\discretionary{/}{%
}{/}S0047404500019849}}


\bibitem{aielloInventorizingSituatingTransforming2020}
G.~Aiello.
\newblock Inventorizing, {{Situating}}, {{Transforming}}: {{Social}} {{Semiotics}} and {{Data}} {{Visualization}}.
\newblock In M.~Engebretsen and H.~Kennedy, eds., {\em Data {{Visualization}} in {{Society}}}, pp. 49--62. Amsterdam University Press, Amsterdam, Netherlands, 2020. \href{https://doi.org/10.5117/9789463722902}
{doi: \textsf{%
10\hspace{.1pt}\discretionary{.}{%
}{.}\hspace{.4pt}5117\discretionary{/}{%
}{/}9789463722902}}


\bibitem{alebriEmbellishmentsRevisitedPerceptions2024}
M.~Alebri, E.~Costanza, G.~Panagiotidou, and D.~P. Brumby.
\newblock Embellishments {{Revisited}}: {{Perceptions}} of {{Embellished Visualisations Through}} the {{Viewer}}'s {{Lens}}.
\newblock {\em IEEE Transactions on Visualization and Computer Graphics}, 30(1), 2024. \href{https://doi.org/10.1109/TVCG.2023.3326914}
{doi: \textsf{%
10\hspace{.1pt}\discretionary{.}{%
}{.}\hspace{.4pt}1109\discretionary{/}{%
}{/}TVCG\hspace{.1pt}\discretionary{.}{%
}{.}\hspace{.4pt}2023\hspace{.1pt}\discretionary{.}{%
}{.}\hspace{.4pt}3326914}}


\bibitem{ashley_tumblr_2019}
V.~Ashley.
\newblock Tumblr {{Porn}} {{Eulogy}}.
\newblock {\em Porn Studies}, 6(3):359--362, July 2019. \href{https://doi.org/10.1080/23268743.2019.1631560}
{doi: \textsf{%
10\hspace{.1pt}\discretionary{.}{%
}{.}\hspace{.4pt}1080\discretionary{/}{%
}{/}23268743\hspace{.1pt}\discretionary{.}{%
}{.}\hspace{.4pt}2019\hspace{.1pt}\discretionary{.}{%
}{.}\hspace{.4pt}1631560}}


\bibitem{bakhtin1981discourse}
M.~Bakhtin.
\newblock {\em The Dialogic Imagination}.
\newblock University of Texas, Austin, TX, USA, 1981.

\bibitem{batemanUsefulJunkEffects2010}
S.~Bateman, R.~L. Mandryk, C.~Gutwin, A.~Genest, D.~McDine, and C.~Brooks.
\newblock Useful {{Junk}}? {{The Effects}} of {{Visual Embellishment}} on {{Comprehension}} and {{Memorability}} of {{Charts}}.
\newblock In {\em Proceedings of the {{SIGCHI Conference}} on {{Human Factors}} in {{Computing Systems}}}, pp. 2573--2582. ACM, New York, USA, 2010. \href{https://doi.org/10.1145/1753326.1753716}
{doi: \textsf{%
10\hspace{.1pt}\discretionary{.}{%
}{.}\hspace{.4pt}1145\discretionary{/}{%
}{/}1753326\hspace{.1pt}\discretionary{.}{%
}{.}\hspace{.4pt}1753716}}


\bibitem{becker1963labeling}
H.~Becker.
\newblock {\em Outsiders: Studies in the Sociology of Deviance}.
\newblock The Free Press of Glencoe, Glencoe, IL, US, 1963.

\bibitem{Bertin1983}
J.~Bertin.
\newblock {\em Semiology of {{Graphics}}: {{Diagrams}}, {{Networks}}, {{Maps}}}.
\newblock University of Wisconsin Press, Madison, WI, 1983.

\bibitem{bilaniuk2003gender}
L.~Bilaniuk.
\newblock Gender, {{Language}} {{Attitudes}}, and {{Language}} {{Status}} in {{Ukraine}}.
\newblock {\em Language in society}, 32(1):47--78, 2003. \href{https://doi.org/10.1017/S0047404503321037}
{doi: \textsf{%
10\hspace{.1pt}\discretionary{.}{%
}{.}\hspace{.4pt}1017\discretionary{/}{%
}{/}S0047404503321037}}


\bibitem{blomSocialDeterminatesVerbal1972}
J.~Blom and J.~Gumperz.
\newblock Some {Social} {Determinates} of {Verbal} {Behavior}.
\newblock In J.~Gumperz and D.~Hymes, eds., {\em Directions in {Sociolinguistics}: {The} {Ethnography} of {Communication}}. Holt Rinehart \& Winston., New York, NY, US, 1972.

\bibitem{borkinMemorabilityVisualizationRecognition2016}
M.~A. Borkin, Z.~Bylinskii, N.~W. Kim, C.~M. Bainbridge, C.~S. Yeh, D.~Borkin, H.~Pfister, and A.~Oliva.
\newblock Beyond {Memorability}: {Visualization} {Recognition} and {Recall}.
\newblock {\em IEEE Transactions on Visualization and Computer Graphics}, 22(1):519--528, 2016. \href{https://doi.org/10.1109/TVCG.2015.2467732}
{doi: \textsf{%
10\hspace{.1pt}\discretionary{.}{%
}{.}\hspace{.4pt}1109\discretionary{/}{%
}{/}TVCG\hspace{.1pt}\discretionary{.}{%
}{.}\hspace{.4pt}2015\hspace{.1pt}\discretionary{.}{%
}{.}\hspace{.4pt}2467732}}


\bibitem{borkinWhatMakesVisualization2013}
M.~A. Borkin, A.~A. Vo, Z.~Bylinskii, P.~Isola, S.~Sunkavalli, A.~Oliva, and H.~Pfister.
\newblock What {Makes} a {Visualization} {Memorable}?
\newblock {\em IEEE Transactions on Visualization and Computer Graphics}, 19(12):2306--2315, Dec. 2013. \href{https://doi.org/10.1109/TVCG.2013.234}
{doi: \textsf{%
10\hspace{.1pt}\discretionary{.}{%
}{.}\hspace{.4pt}1109\discretionary{/}{%
}{/}TVCG\hspace{.1pt}\discretionary{.}{%
}{.}\hspace{.4pt}2013\hspace{.1pt}\discretionary{.}{%
}{.}\hspace{.4pt}234}}


\bibitem{bourhis1976language}
R.~Y. Bourhis and H.~Giles.
\newblock The {{Language}} of {{Coperation}} in {{Wales}}: {{A}} {{Field}} {{Study}}.
\newblock {\em Language sciences}, 42(13-16), 1976.

\bibitem{CardCHI11999}
S.~K. Card, J.~D. Mackinlay, and B.~Shneiderman.
\newblock {\em Readings in {Information} {Visualization}: {Using} {Vision} to {Think}}.
\newblock Academic Press, San Diego, CA, US, 1999.

\bibitem{carrie2018american}
E.~Carrie and R.~M. McKenzie.
\newblock American or british? l2 speakers’ recognition and evaluations of accent features in english.
\newblock {\em Journal of Multilingual and Multicultural Development}, 39(4):313--328, 2018. \href{https://doi.org/10.1080/01434632.2017.1389946}
{doi: \textsf{%
10\hspace{.1pt}\discretionary{.}{%
}{.}\hspace{.4pt}1080\discretionary{/}{%
}{/}01434632\hspace{.1pt}\discretionary{.}{%
}{.}\hspace{.4pt}2017\hspace{.1pt}\discretionary{.}{%
}{.}\hspace{.4pt}1389946}}


\bibitem{charmazGroundedTheoryEthnography2001}
K.~Charmaz and R.~G. Mitchell.
\newblock Grounded {Theory} in {Ethnography}.
\newblock In {\em Handbook of {Ethnography}}, pp. 160--174. SAGE Publications Ltd, 2001.

\bibitem{choski2021script}
N.~Choski.
\newblock {\em Graphic Politics in Eastern India: Script and the Quest for Autonomy}.
\newblock Bloomsbury Publishing, London, UK, 2021.

\bibitem{dame_making_2016}
A.~Dame.
\newblock Making a name for yourself: tagging as transgender ontological practice on {Tumblr}.
\newblock {\em Critical Studies in Media Communication}, Jan. 2016. \href{https://doi.org/10.1080/15295036.2015.1130846}
{doi: \textsf{%
10\hspace{.1pt}\discretionary{.}{%
}{.}\hspace{.4pt}1080\discretionary{/}{%
}{/}15295036\hspace{.1pt}\discretionary{.}{%
}{.}\hspace{.4pt}2015\hspace{.1pt}\discretionary{.}{%
}{.}\hspace{.4pt}1130846}}


\bibitem{donzelli2024chartsJLA}
A.~Donzelli.
\newblock From mandela to flowchart: Managerial governmentality and the evidentiary technologies of indonesia's reformasi.
\newblock {\em Journal of Linguistic Anthropology}, 34(2):290--319, 2024. \href{https://doi.org/10.1111/jola.12435}
{doi: \textsf{%
10\hspace{.1pt}\discretionary{.}{%
}{.}\hspace{.4pt}1111\discretionary{/}{%
}{/}jola\hspace{.1pt}\discretionary{.}{%
}{.}\hspace{.4pt}12435}}


\bibitem{dourishReadingInterpretingEthnography2014a}
P.~Dourish.
\newblock Reading and {{Interpreting Ethnography}}.
\newblock In {\em Ways of {{Knowing}} in {{HCI}}}, pp. 1--23. Springer, NY, 2014. \href{https://doi.org/10.1007/978-1-4939-0378-8_1}
{doi: \textsf{%
10\hspace{.1pt}\discretionary{.}{%
}{.}\hspace{.4pt}1007\discretionary{/}{%
}{/}978\discretionary{%
}{-}{-}1\discretionary{%
}{-}{-}4939\discretionary{%
}{-}{-}0378\discretionary{%
}{-}{-}8\_1}}


\bibitem{durantigoodwin1992context}
A.~Duranti and C.~Goodwin.
\newblock {\em Rethinking context: Language as an interactive phenomenon}.
\newblock Cambridge University Press, 1992.

\bibitem{eckert1999beltenhigh}
P.~Eckert.
\newblock {\em Language Variation as Social Practice: The Linguistic Construction of Identity in Belten High}.
\newblock Blackwell, 1999.

\bibitem{eckertVariationIndexicalField2008}
P.~Eckert.
\newblock Variation and the indexical field 1.
\newblock {\em Journal of Sociolinguistics}, 12:453--476, Jan 2008. \href{https://doi.org/10.1111/j.1467-9841.2008.00374.x}
{doi: \textsf{%
10\hspace{.1pt}\discretionary{.}{%
}{.}\hspace{.4pt}1111\discretionary{/}{%
}{/}j\hspace{.1pt}\discretionary{.}{%
}{.}\hspace{.4pt}1467\discretionary{%
}{-}{-}9841\hspace{.1pt}\discretionary{.}{%
}{.}\hspace{.4pt}2008\hspace{.1pt}\discretionary{.}{%
}{.}\hspace{.4pt}00374\hspace{.1pt}\discretionary{.}{%
}{.}\hspace{.4pt}x}}


\bibitem{elhamdadiHowWeMeasure2022a}
H.~Elhamdadi, A.~Gaba, Y.-S. Kim, and C.~Xiong.
\newblock How do we measure trust in visual data communication?
\newblock In {\em 2022 {IEEE} {Evaluation} and {Beyond}-{Methodological} {Approaches} for {Visualization} ({BELIV})}. IEEE, 2022. \href{https://doi.org/10.1109/BELIV57783.2022.00014}
{doi: \textsf{%
10\hspace{.1pt}\discretionary{.}{%
}{.}\hspace{.4pt}1109\discretionary{/}{%
}{/}BELIV57783\hspace{.1pt}\discretionary{.}{%
}{.}\hspace{.4pt}2022\hspace{.1pt}\discretionary{.}{%
}{.}\hspace{.4pt}00014}}


\bibitem{elhamdadiVistrustMultidimensionalFramework2023a}
H.~Elhamdadi, A.~Stefkovics, J.~Beyer, E.~Moerth, H.~Pfister, C.~X. Bearfield, and C.~Nobre.
\newblock Vistrust: a {Multidimensional} {Framework} and {Empirical} {Study} of {Trust} in {Data} {Visualizations}.
\newblock {\em IEEE Transactions on Visualization and Computer Graphics}, pp. 1--11, 2023. \href{https://doi.org/10.1109/TVCG.2023.3326579}
{doi: \textsf{%
10\hspace{.1pt}\discretionary{.}{%
}{.}\hspace{.4pt}1109\discretionary{/}{%
}{/}TVCG\hspace{.1pt}\discretionary{.}{%
}{.}\hspace{.4pt}2023\hspace{.1pt}\discretionary{.}{%
}{.}\hspace{.4pt}3326579}}


\bibitem{Fox_VisPsych_theoriesmodels_2023}
A.~Fox.
\newblock Theories and {Models} of {Graph} {Comprehension}.
\newblock In M.~Chen, D.~A. Szafir, R.~Borgo, L.~M.~K. Padilla, D.~J. Edwards, and B.~Fisher, eds., {\em Visualization {Psychology}}. Springer, 2023. \href{https://doi.org/10.1007/978-3-031-34738-2_2}
{doi: \textsf{%
10\hspace{.1pt}\discretionary{.}{%
}{.}\hspace{.4pt}1007\discretionary{/}{%
}{/}978\discretionary{%
}{-}{-}3\discretionary{%
}{-}{-}031\discretionary{%
}{-}{-}34738\discretionary{%
}{-}{-}2\_2}}


\bibitem{FoxHollan_VisPsych_researchprogramme2023}
A.~Fox and J.~D. Hollan.
\newblock Visualization {Psychology}: {Foundations} for an {Interdisciplinary} {Research} {Programme}.
\newblock In M.~Chen, D.~A. Szafir, R.~Borgo, L.~M.~K. Padilla, D.~J. Edwards, and B.~Fisher, eds., {\em Visualization {Psychology}}. Springer, 2023. \href{https://doi.org/10.1007/978-3-031-34738-2_9}
{doi: \textsf{%
10\hspace{.1pt}\discretionary{.}{%
}{.}\hspace{.4pt}1007\discretionary{/}{%
}{/}978\discretionary{%
}{-}{-}3\discretionary{%
}{-}{-}031\discretionary{%
}{-}{-}34738\discretionary{%
}{-}{-}2\_9}}


\bibitem{PAPER2}
A.~Fox, M.~Morgenstern, G.~Jones, and A.~Satyanarayan.
\newblock Quantifying {{Visualization}} {{Vibes}}: {{Measuring}} {{Socio-Indexicality}} at {{Scale}}.
\newblock {\em IEEE Transactions on Visualization and Computer Graphics}, 32(1), 2026.

\bibitem{franconeriScienceVisualData2021}
S.~L. Franconeri, L.~M. Padilla, P.~Shah, J.~M. Zacks, and J.~Hullman.
\newblock The {{Science}} of {{Visual Data Communication}}: {{What Works}}.
\newblock {\em Psychological Science in the Public Interest}, 22(3):110--161, Dec. 2021. \href{https://doi.org/10.1177/15291006211051956}
{doi: \textsf{%
10\hspace{.1pt}\discretionary{.}{%
}{.}\hspace{.4pt}1177\discretionary{/}{%
}{/}15291006211051956}}


\bibitem{friggModelsScience2006}
R.~Frigg and S.~Hartmann.
\newblock {Models in Science}.
\newblock In E.~N. Zalta and U.~Nodelman, eds., {\em The {Stanford} Encyclopedia of Philosophy}. Metaphysics Research Lab, Stanford University, {S}ummer 2025 ed., 2025.

\bibitem{gal2015politics}
S.~Gal.
\newblock Politics of translation.
\newblock {\em Annual Review of Anthropology}, 44(1):225--240, 2015. \href{https://doi.org/10.1146/annurev-anthro-102214-013806}
{doi: \textsf{%
10\hspace{.1pt}\discretionary{.}{%
}{.}\hspace{.4pt}1146\discretionary{/}{%
}{/}annurev\discretionary{%
}{-}{-}anthro\discretionary{%
}{-}{-}102214\discretionary{%
}{-}{-}013806}}


\bibitem{garrett2010attitudes}
P.~Garrett.
\newblock {\em Attitudes to language}.
\newblock Cambridge University Press, 2010.

\bibitem{gilesbillings2004attitude}
A.~Giles, H. \&~Billings.
\newblock Assessing language attitudes: Speaker evaluation studies.
\newblock In {\em The Handbook of Applied Linguistics}. Blackwell, 2004.

\bibitem{goffmanFormsTalk1981a}
E.~Goffman.
\newblock {\em Forms of {Talk}}.
\newblock University of Pennsylvania Press, Philadelphia, PA, UA, 1981.

\bibitem{hallIdentityInteractionSociocultural2005}
K.~Hall and M.~Bucholtz.
\newblock Identity and {Interaction}: {A} {Sociocultural} {Linguistic} {Approach}.
\newblock {\em Discourse Studies}, 7(4-5):585--614, 2005.
\newblock Publisher: Sage Publications. \href{https://doi.org/10.1177/1461445605054407}
{doi: \textsf{%
10\hspace{.1pt}\discretionary{.}{%
}{.}\hspace{.4pt}1177\discretionary{/}{%
}{/}1461445605054407}}


\bibitem{harrisonInfographicAestheticsDesigning2015}
L.~Harrison, K.~Reinecke, and R.~Chang.
\newblock Infographic {{Aesthetics}}: {{Designing}} for the {{First Impression}}.
\newblock In {\em Proceedings of the 33rd {{Annual ACM Conference}} on {{Human Factors}} in {{Computing Systems}}}, pp. 1187--1190. ACM, New York, USA, 2015. \href{https://doi.org/10.1145/2702123.2702545}
{doi: \textsf{%
10\hspace{.1pt}\discretionary{.}{%
}{.}\hspace{.4pt}1145\discretionary{/}{%
}{/}2702123\hspace{.1pt}\discretionary{.}{%
}{.}\hspace{.4pt}2702545}}


\bibitem{haugh_revisiting_2004}
M.~Haugh.
\newblock Revisiting the conceptualisation of politeness in {English} and {Japanese}.
\newblock {\em Multilingua}, 23(1-2):85--109, Mar. 2004.
\newblock Publisher: De Gruyter Mouton Section: Multilingua. \href{https://doi.org/10.1515/mult.2004.009}
{doi: \textsf{%
10\hspace{.1pt}\discretionary{.}{%
}{.}\hspace{.4pt}1515\discretionary{/}{%
}{/}mult\hspace{.1pt}\discretionary{.}{%
}{.}\hspace{.4pt}2004\hspace{.1pt}\discretionary{.}{%
}{.}\hspace{.4pt}009}}


\bibitem{heEnthusiasticGroundedAvoidant2024}
H.~A. He, J.~Walny, S.~Thoma, S.~Carpendale, and W.~Willett.
\newblock Enthusiastic and {Grounded}, {Avoidant} and {Cautious}: {Understanding} {Public} {Receptivity} to {Data} and {Visualizations}.
\newblock {\em IEEE Transactions on Visualization and Computer Graphics}, 30(1), 2024. \href{https://doi.org/10.1109/TVCG.2023.3326917}
{doi: \textsf{%
10\hspace{.1pt}\discretionary{.}{%
}{.}\hspace{.4pt}1109\discretionary{/}{%
}{/}TVCG\hspace{.1pt}\discretionary{.}{%
}{.}\hspace{.4pt}2023\hspace{.1pt}\discretionary{.}{%
}{.}\hspace{.4pt}3326917}}


\bibitem{hoganElicitationInterviewTechnique2016}
T.~Hogan, U.~Hinrichs, and E.~Hornecker.
\newblock The {Elicitation} {Interview} {Technique}: {Capturing} {People}'s {Experiences} of {Data} {Representations}.
\newblock {\em IEEE Transactions on Visualization and Computer Graphics}, 22(12):2579--2593, Dec. 2016. \href{https://doi.org/10.1109/TVCG.2015.2511718}
{doi: \textsf{%
10\hspace{.1pt}\discretionary{.}{%
}{.}\hspace{.4pt}1109\discretionary{/}{%
}{/}TVCG\hspace{.1pt}\discretionary{.}{%
}{.}\hspace{.4pt}2015\hspace{.1pt}\discretionary{.}{%
}{.}\hspace{.4pt}2511718}}


\bibitem{hullmanVisualizationRhetoricFraming2011b}
J.~Hullman and N.~Diakopoulos.
\newblock Visualization {Rhetoric}: {Framing} {Effects} in {Narrative} {Visualization}.
\newblock {\em IEEE Transactions on Visualization and Computer Graphics}, 17(12), Dec. 2011. \href{https://doi.org/10.1109/TVCG.2011.255}
{doi: \textsf{%
10\hspace{.1pt}\discretionary{.}{%
}{.}\hspace{.4pt}1109\discretionary{/}{%
}{/}TVCG\hspace{.1pt}\discretionary{.}{%
}{.}\hspace{.4pt}2011\hspace{.1pt}\discretionary{.}{%
}{.}\hspace{.4pt}255}}


\bibitem{hullmanContentContextCritique2015}
J.~Hullman, N.~Diakopoulos, E.~Momeni, and E.~Adar.
\newblock Content, {Context}, and {Critique}: {Commenting} on a {Data} {Visualization} {Blog}.
\newblock In {\em Proceedings of the 18th {ACM} {Conference} on {Computer} {Supported} {Cooperative} {Work} \& {Social} {Computing}}, {CSCW} '15, pp. 1170--1175. ACM, New York, USA, Feb. 2015. \href{https://doi.org/10.1145/2675133.2675207}
{doi: \textsf{%
10\hspace{.1pt}\discretionary{.}{%
}{.}\hspace{.4pt}1145\discretionary{/}{%
}{/}2675133\hspace{.1pt}\discretionary{.}{%
}{.}\hspace{.4pt}2675207}}


\bibitem{irvine2001style}
J.~Irvine.
\newblock ‘style’ as distinctiveness: The culture and ideology of linguistic differentiation.
\newblock In {\em Style and Sociolinguistic Variation}. Cambridge University Press, New York, NY, US, 2004.

\bibitem{jabareenBuildingConceptualFramework2009}
Y.~Jabareen.
\newblock Building a {{Conceptual Framework}}: {{Philosophy}}, {{Definitions}}, and {{Procedure}}.
\newblock {\em International Journal of Qualitative Methods}, 8(4):49--62. \href{https://doi.org/10.1177/160940690900800406}
{doi: \textsf{%
10\hspace{.1pt}\discretionary{.}{%
}{.}\hspace{.4pt}1177\discretionary{/}{%
}{/}160940690900800406}}


\bibitem{jakobson1960poetics}
R.~Jakobson.
\newblock Linguistics and poetics.
\newblock In {\em Style in Language}. MIT Press, Cambridge, MA, US, 1960.

\bibitem{johnstone2008indexicality}
B.~Johnstone and S.~F. Kiesling.
\newblock Indexicality and experience: Exploring the meanings of /aw/ -monophthongization in {P}ittsburgh.
\newblock {\em Journal of {Socio- linguistics}}, 12(1):5--33, 2008. \href{https://doi.org/10.1111/j.1467-9841.2008.00351.x}
{doi: \textsf{%
10\hspace{.1pt}\discretionary{.}{%
}{.}\hspace{.4pt}1111\discretionary{/}{%
}{/}j\hspace{.1pt}\discretionary{.}{%
}{.}\hspace{.4pt}1467\discretionary{%
}{-}{-}9841\hspace{.1pt}\discretionary{.}{%
}{.}\hspace{.4pt}2008\hspace{.1pt}\discretionary{.}{%
}{.}\hspace{.4pt}00351\hspace{.1pt}\discretionary{.}{%
}{.}\hspace{.4pt}x}}


\bibitem{jones2009enquoting}
G.~M. Jones and B.~B. Schieffelin.
\newblock Enquoting voices, accomplishing talk: Uses of be+ like in instant messaging.
\newblock {\em Language \& Communication}, 29(1):77--113, 2009. \href{https://doi.org/10.1016/j.langcom.2007.09.003}
{doi: \textsf{%
10\hspace{.1pt}\discretionary{.}{%
}{.}\hspace{.4pt}1016\discretionary{/}{%
}{/}j\hspace{.1pt}\discretionary{.}{%
}{.}\hspace{.4pt}langcom\hspace{.1pt}\discretionary{.}{%
}{.}\hspace{.4pt}2007\hspace{.1pt}\discretionary{.}{%
}{.}\hspace{.4pt}09\hspace{.1pt}\discretionary{.}{%
}{.}\hspace{.4pt}003}}


\bibitem{jones2019testifyingblack}
T.~Jones, J.~Rose~Kalbfeld, R.~Hancock, and R.~Clark.
\newblock Testifying while black: An experimental study of cohort reporter accuracy in transcription of african american english.
\newblock {\em Language}, 95(2):216--252, 2019. \href{https://doi.org/10.1353/lan.2019.0042}
{doi: \textsf{%
10\hspace{.1pt}\discretionary{.}{%
}{.}\hspace{.4pt}1353\discretionary{/}{%
}{/}lan\hspace{.1pt}\discretionary{.}{%
}{.}\hspace{.4pt}2019\hspace{.1pt}\discretionary{.}{%
}{.}\hspace{.4pt}0042}}


\bibitem{kauerPublicLifeData2021a}
T.~Kauer, M.~Dörk, A.~L. Ridley, and B.~Bach.
\newblock The {{Public}} {{Life}} of {{Data}}: {{Investigating}} {{Reactions}} to {{Visualizations}} on {{Reddit}}.
\newblock In {\em Proceedings of the 2021 {CHI} {Conference} on {Human} {Factors} in {Computing} {Systems}}, pp. 1--12. ACM, New York, USA, May 2021. \href{https://doi.org/10.1145/3411764.3445720}
{doi: \textsf{%
10\hspace{.1pt}\discretionary{.}{%
}{.}\hspace{.4pt}1145\discretionary{/}{%
}{/}3411764\hspace{.1pt}\discretionary{.}{%
}{.}\hspace{.4pt}3445720}}


\bibitem{keaneSemioticIdeology2018}
W.~Keane.
\newblock On {Semiotic} {Ideology}.
\newblock {\em Signs and Society}, 6(1):64--87, Jan. 2018.
\newblock Publisher: The University of Chicago Press. \href{https://doi.org/10.1086/695387}
{doi: \textsf{%
10\hspace{.1pt}\discretionary{.}{%
}{.}\hspace{.4pt}1086\discretionary{/}{%
}{/}695387}}


\bibitem{keatingWordsMatterCommunicating2016}
E.~Keating and S.~L. Jarvenpaa.
\newblock {\em Words {Matter}: {Communicating} {Effectively} in the {New} {Global} {Office}}.
\newblock University of California Press, 2016.

\bibitem{kennedyEngagingBigData2016}
H.~Kennedy, R.~L. Hill, W.~Allen, and A.~Kirk.
\newblock Engaging with (big) data visualizations: {Factors} that affect engagement and resulting new definitions of effectiveness.
\newblock {\em First Monday}, Nov. 2016. \href{https://doi.org/10.5210/fm.v21i11.6389}
{doi: \textsf{%
10\hspace{.1pt}\discretionary{.}{%
}{.}\hspace{.4pt}5210\discretionary{/}{%
}{/}fm\hspace{.1pt}\discretionary{.}{%
}{.}\hspace{.4pt}v21i11\hspace{.1pt}\discretionary{.}{%
}{.}\hspace{.4pt}6389}}


\bibitem{kosara2013storytelling}
R.~Kosara and J.~Mackinlay.
\newblock Storytelling: The next step for visualization.
\newblock {\em Computer}, 46(5):44--50, 2013. \href{https://doi.org/10.1109/MC.2013.36}
{doi: \textsf{%
10\hspace{.1pt}\discretionary{.}{%
}{.}\hspace{.4pt}1109\discretionary{/}{%
}{/}MC\hspace{.1pt}\discretionary{.}{%
}{.}\hspace{.4pt}2013\hspace{.1pt}\discretionary{.}{%
}{.}\hspace{.4pt}36}}


\bibitem{kostelnickVisualRhetoricData2007}
C.~Kostelnick.
\newblock The {{Visual Rhetoric}} of {{Data Displays}}: {{The Conundrum}} of {{Clarity}}.
\newblock {\em IEEE Transactions on Professional Communication}, 50(4):280--294, Dec. 2007. \href{https://doi.org/10.1109/TPC.2007.908725}
{doi: \textsf{%
10\hspace{.1pt}\discretionary{.}{%
}{.}\hspace{.4pt}1109\discretionary{/}{%
}{/}TPC\hspace{.1pt}\discretionary{.}{%
}{.}\hspace{.4pt}2007\hspace{.1pt}\discretionary{.}{%
}{.}\hspace{.4pt}908725}}


\bibitem{ladegaard2000language}
H.~J. Ladegaard.
\newblock Language attitudes and sociolinguistic behaviour: Exploring attitude-behaviour relations in language.
\newblock {\em Journal of sociolinguistics}, 4(2):214--233, 2000. \href{https://doi.org/10.1111/1467-9481.00112}
{doi: \textsf{%
10\hspace{.1pt}\discretionary{.}{%
}{.}\hspace{.4pt}1111\discretionary{/}{%
}{/}1467\discretionary{%
}{-}{-}9481\hspace{.1pt}\discretionary{.}{%
}{.}\hspace{.4pt}00112}}


\bibitem{lambert1960evaluational}
W.~E. Lambert, R.~C. Hodgson, R.~C. Gardner, and S.~Fillenbaum.
\newblock Evaluational reactions to spoken languages.
\newblock {\em The journal of abnormal and social psychology}, 60(1):44, 1960.

\bibitem{lanAffectiveVisualizationDesign2024a}
X.~Lan, Y.~Wu, and N.~Cao.
\newblock Affective {Visualization} {Design}: {Leveraging} the {Emotional} {Impact} of {Data}.
\newblock {\em IEEE Transactions on Visualization and Computer Graphics}, 30(1), Jan. 2024. \href{https://doi.org/10.1109/TVCG.2023.3327385}
{doi: \textsf{%
10\hspace{.1pt}\discretionary{.}{%
}{.}\hspace{.4pt}1109\discretionary{/}{%
}{/}TVCG\hspace{.1pt}\discretionary{.}{%
}{.}\hspace{.4pt}2023\hspace{.1pt}\discretionary{.}{%
}{.}\hspace{.4pt}3327385}}


\bibitem{latour1987action}
B.~Latour.
\newblock {\em Science in Action: How to Follow Scientists and Engineers through Society}.
\newblock Harvard University Press, Cambridge, MA, US, 1987.

\bibitem{lauModelInformationAesthetics2007}
A.~Lau and A.~Vande~Moere.
\newblock Towards a {{Model}} of {{Information Aesthetics}} in {{Information Visualization}}.
\newblock {\em International Conference Information Visualization}, pp. 87--92, 2007. \href{https://doi.org/10.1109/IV.2007.114}
{doi: \textsf{%
10\hspace{.1pt}\discretionary{.}{%
}{.}\hspace{.4pt}1109\discretionary{/}{%
}{/}IV\hspace{.1pt}\discretionary{.}{%
}{.}\hspace{.4pt}2007\hspace{.1pt}\discretionary{.}{%
}{.}\hspace{.4pt}114}}


\bibitem{Leder2004}
H.~Leder, B.~Belke, A.~Oeberst, and D.~Augustin.
\newblock A model of aesthetic appreciation and aesthetic judgments.
\newblock {\em British journal of psychology}, 95, 2004. \href{https://doi.org/10.1348/0007126042369811}
{doi: \textsf{%
10\hspace{.1pt}\discretionary{.}{%
}{.}\hspace{.4pt}1348\discretionary{/}{%
}{/}0007126042369811}}


\bibitem{leeViralVisualizationsHow2021}
C.~Lee, T.~Yang, G.~Inchoco, G.~M. Jones, and A.~Satyanarayan.
\newblock Viral {Visualizations}: {How} {Coronavirus} {Skeptics} {Use} {Orthodox} {Data} {Practices} to {Promote} {Unorthodox} {Science} {Online}.
\newblock In {\em {ACM} {Human} {Factors} in {Computing} {Systems} ({CHI})}. May 2021. \href{https://doi.org/10.1145/3411764.3445211}
{doi: \textsf{%
10\hspace{.1pt}\discretionary{.}{%
}{.}\hspace{.4pt}1145\discretionary{/}{%
}{/}3411764\hspace{.1pt}\discretionary{.}{%
}{.}\hspace{.4pt}3445211}}


\bibitem{lee-robbinsAffectiveLearningObjectives2023a}
E.~Lee-Robbins and E.~Adar.
\newblock Affective {Learning} {Objectives} for {Communicative} {Visualizations}.
\newblock {\em IEEE Transactions on Visualization and Computer Graphics}, 29(01):1--11, Jan. 2023. \href{https://doi.org/10.1109/TVCG.2022.3209500}
{doi: \textsf{%
10\hspace{.1pt}\discretionary{.}{%
}{.}\hspace{.4pt}1109\discretionary{/}{%
}{/}TVCG\hspace{.1pt}\discretionary{.}{%
}{.}\hspace{.4pt}2022\hspace{.1pt}\discretionary{.}{%
}{.}\hspace{.4pt}3209500}}


\bibitem{levyPersonCenteredInterviewingObservation2015}
R.~Levy and D.~Hollan.
\newblock Person-{Centered} {Interviewing} and {Observation}.
\newblock In {\em Handbook of {Methods} in {Cultural} {Anthropology}}, pp. 313--342. Rowman \& Littlefield, London, UK, 2015.

\bibitem{linVisualizationAestheticsBias2021a}
C.~Lin and M.~A. Thornton.
\newblock Visualization aesthetics bias trust in science, news, and social media, Dec. 2021. \href{https://doi.org/10.31234/osf.io/dnr9s}
{doi: \textsf{%
10\hspace{.1pt}\discretionary{.}{%
}{.}\hspace{.4pt}31234\discretionary{/}{%
}{/}osf\hspace{.1pt}\discretionary{.}{%
}{.}\hspace{.4pt}io\discretionary{/}{%
}{/}dnr9s}}


\bibitem{lisnicMisleadingVisualTricks2023}
M.~Lisnic, C.~Polychronis, A.~Lex, and M.~Kogan.
\newblock Misleading {{Beyond Visual Tricks}}: {{How People Actually Lie}} with {{Charts}}.
\newblock In {\em Proceedings of the 2023 {{CHI Conference}} on {{Human Factors}} in {{Computing Systems}}}, pp. 1--21. ACM. \href{https://doi.org/10.1145/3544548.3580910}
{doi: \textsf{%
10\hspace{.1pt}\discretionary{.}{%
}{.}\hspace{.4pt}1145\discretionary{/}{%
}{/}3544548\hspace{.1pt}\discretionary{.}{%
}{.}\hspace{.4pt}3580910}}


\bibitem{mahyarTaxonomyEvaluatingUser2015}
N.~Mahyar, S.-H. Kim, and B.~C. Kwon.
\newblock Towards a taxonomy for evaluating user engagement in information visualization.
\newblock In {\em Workshop on {{Personal Visualization}}: {{Exploring Everyday Life}}}, vol.~3, p.~4, 2015.

\bibitem{markantCanDataVisualizations2022a}
D.~B. Markant, M.~Rogha, A.~Karduni, R.~Wesslen, and W.~Dou.
\newblock Can {Data} {Visualizations} {Change} {Minds}? {Identifying} {Mechanisms} of {Elaborative} {Thinking} and {Persuasion}.
\newblock In {\em 2022 {IEEE} {Workshop} on {Visualization} for {Social} {Good}}, pp. 1--5, Oct. 2022. \href{https://doi.org/10.1109/VIS4Good57762.2022.00005}
{doi: \textsf{%
10\hspace{.1pt}\discretionary{.}{%
}{.}\hspace{.4pt}1109\discretionary{/}{%
}{/}VIS4Good57762\hspace{.1pt}\discretionary{.}{%
}{.}\hspace{.4pt}2022\hspace{.1pt}\discretionary{.}{%
}{.}\hspace{.4pt}00005}}


\bibitem{massoudObjectificationInscriptionKnowledge2008}
L.~A. Massoud and J.~C. Kuipers.
\newblock Objectification and the inscription of knowledge in science classrooms.
\newblock {\em Linguistics and Education}, 19(3):211--224, Dec. 2008. \href{https://doi.org/10.1016/j.linged.2008.05.009}
{doi: \textsf{%
10\hspace{.1pt}\discretionary{.}{%
}{.}\hspace{.4pt}1016\discretionary{/}{%
}{/}j\hspace{.1pt}\discretionary{.}{%
}{.}\hspace{.4pt}linged\hspace{.1pt}\discretionary{.}{%
}{.}\hspace{.4pt}2008\hspace{.1pt}\discretionary{.}{%
}{.}\hspace{.4pt}05\hspace{.1pt}\discretionary{.}{%
}{.}\hspace{.4pt}009}}


\bibitem{mccracken_tumblr_2017}
A.~McCracken.
\newblock Tumblr {Youth} {Subcultures} and {Media} {Engagement}.
\newblock {\em Cinema Journal}, 57(1):151--161, 2017. \href{https://doi.org/10.1353/cj.2017.0061}
{doi: \textsf{%
10\hspace{.1pt}\discretionary{.}{%
}{.}\hspace{.4pt}1353\discretionary{/}{%
}{/}cj\hspace{.1pt}\discretionary{.}{%
}{.}\hspace{.4pt}2017\hspace{.1pt}\discretionary{.}{%
}{.}\hspace{.4pt}0061}}


\bibitem{mendoza-dentonLanguageIdentity2004}
N.~Mendoza‐Denton.
\newblock Language and {Identity}.
\newblock In J.~K. Chambers, P.~Trudgill, and N.~Schilling‐Estes, eds., {\em The {Handbook} of {Language} {Variation} and {Change}}, pp. 475--499. Wiley, 2004. \href{https://doi.org/10.1002/9780470756591.ch19}
{doi: \textsf{%
10\hspace{.1pt}\discretionary{.}{%
}{.}\hspace{.4pt}1002\discretionary{/}{%
}{/}9780470756591\hspace{.1pt}\discretionary{.}{%
}{.}\hspace{.4pt}ch19}}


\bibitem{moereRoleDesignInformation2011}
A.~V. Moere and H.~Purchase.
\newblock On the role of design in information visualization.
\newblock {\em Information Visualization}, 10(4):356--371, Oct. 2011. \href{https://doi.org/10.1177/1473871611415996}
{doi: \textsf{%
10\hspace{.1pt}\discretionary{.}{%
}{.}\hspace{.4pt}1177\discretionary{/}{%
}{/}1473871611415996}}


\bibitem{michelle-diss}
M.~Morgenstern.
\newblock {\em Play, Purity, and the Poetics of Moral Counterpublic Discourse on Tumblr.com}.
\newblock PhD thesis, University of Virginia, 2022.

\bibitem{murphyFontroversyHowCare2017}
K.~Murphy.
\newblock Fontroversy! or, how to care about the shape of language.
\newblock In {\em Language and {Materiality}: {Ethnographic} and {Theoretical} {Explorations}}, pp. 63--86. Jan. 2017. \href{https://doi.org/10.1017/9781316848418.004}
{doi: \textsf{%
10\hspace{.1pt}\discretionary{.}{%
}{.}\hspace{.4pt}1017\discretionary{/}{%
}{/}9781316848418\hspace{.1pt}\discretionary{.}{%
}{.}\hspace{.4pt}004}}


\bibitem{murphyFakeNewsWeb2023a}
K.~M. Murphy.
\newblock Fake {News} and the {Web} of {Plausibility}.
\newblock {\em Social Media + Society}, 9(2), Apr. 2023. \href{https://doi.org/10.1177/20563051231170606}
{doi: \textsf{%
10\hspace{.1pt}\discretionary{.}{%
}{.}\hspace{.4pt}1177\discretionary{/}{%
}{/}20563051231170606}}


\bibitem{nakassisLinguisticAnthropology20152016}
C.~V. Nakassis.
\newblock Linguistic {Anthropology} in 2015: {Not} the {Study} of {Language}.
\newblock {\em American Anthropologist}, 118(2), 2016. \href{https://doi.org/10.1111/aman.12528}
{doi: \textsf{%
10\hspace{.1pt}\discretionary{.}{%
}{.}\hspace{.4pt}1111\discretionary{/}{%
}{/}aman\hspace{.1pt}\discretionary{.}{%
}{.}\hspace{.4pt}12528}}


\bibitem{nakassisLinguisticAnthropologyImages2023}
C.~V. Nakassis.
\newblock A {Linguistic} {Anthropology} of {Images}.
\newblock {\em Annual Review of Anthropology}, 52(1):73--91, Oct. 2023. \href{https://doi.org/10.1146/annurev-anthro-052721-092147}
{doi: \textsf{%
10\hspace{.1pt}\discretionary{.}{%
}{.}\hspace{.4pt}1146\discretionary{/}{%
}{/}annurev\discretionary{%
}{-}{-}anthro\discretionary{%
}{-}{-}052721\discretionary{%
}{-}{-}092147}}


\bibitem{obrienStandardsReportingQualitative2014}
B.~C. O'Brien, I.~B. Harris, T.~J. Beckman, D.~A. Reed, and D.~A. Cook.
\newblock Standards for {{Reporting Qualitative Research}}: {{A Synthesis}} of {{Recommendations}}.
\newblock {\em Academic Medicine}, 89(9):1245--1251, Sept. 2014. \href{https://doi.org/10.1097/ACM.0000000000000388}
{doi: \textsf{%
10\hspace{.1pt}\discretionary{.}{%
}{.}\hspace{.4pt}1097\discretionary{/}{%
}{/}ACM\hspace{.1pt}\discretionary{.}{%
}{.}\hspace{.4pt}0000000000000388}}


\bibitem{peckDataPersonalAttitudes2019a}
E.~M. Peck, S.~E. Ayuso, and O.~El-Etr.
\newblock Data is {Personal}: {Attitudes} and {Perceptions} of {Data} {Visualization} in {Rural} {Pennsylvania}.
\newblock In {\em Proceedings of the 2019 {CHI} {Conference} on {Human} {Factors} in {Computing} {Systems}}. ACM, NY, USA, 2019. \href{https://doi.org/10.1145/3290605.3300474}
{doi: \textsf{%
10\hspace{.1pt}\discretionary{.}{%
}{.}\hspace{.4pt}1145\discretionary{/}{%
}{/}3290605\hspace{.1pt}\discretionary{.}{%
}{.}\hspace{.4pt}3300474}}


\bibitem{peirce-hoopes}
C.~S. Peirce.
\newblock {\em Peirce on {Signs}: {Writings} on {Semiotic} by {Charles} {Sanders} {Peirce}}.
\newblock James {Hoopes}. University of North Carolina Press, 1991.

\bibitem{reddyConduitMetaphorCase1993}
M.~J. Reddy.
\newblock The conduit metaphor: {A} case of frame conflict in our language about language.
\newblock In A.~Ortony, ed., {\em Metaphor and {Thought}}, pp. 164--201. Cambridge University Press, 1993. \href{https://doi.org/10.1017/CBO9781139173865.012}
{doi: \textsf{%
10\hspace{.1pt}\discretionary{.}{%
}{.}\hspace{.4pt}1017\discretionary{/}{%
}{/}CBO9781139173865\hspace{.1pt}\discretionary{.}{%
}{.}\hspace{.4pt}012}}


\bibitem{renninger_where_2015}
B.~J. Renninger.
\newblock “{Where} {I} can be myself … where {I} can speak my mind” : {Networked} counterpublics in a polymedia environment.
\newblock {\em New Media \& Society}, 17(9):1513--1529, Oct. 2015. \href{https://doi.org/10.1177/1461444814530095}
{doi: \textsf{%
10\hspace{.1pt}\discretionary{.}{%
}{.}\hspace{.4pt}1177\discretionary{/}{%
}{/}1461444814530095}}


\bibitem{rettbergWaysKnowingData2020}
J.~W. Rettberg.
\newblock Ways of knowing with data visualizations.
\newblock In M.~Engebretsen and H.~Kennedy, eds., {\em Data {{Visualization}} in {{Society}}}, pp. 35--48. Amsterdam University Press, 2020. \href{https://doi.org/10.2307/j.ctvzgb8c7.8}
{doi: \textsf{%
10\hspace{.1pt}\discretionary{.}{%
}{.}\hspace{.4pt}2307\discretionary{/}{%
}{/}j\hspace{.1pt}\discretionary{.}{%
}{.}\hspace{.4pt}ctvzgb8c7\hspace{.1pt}\discretionary{.}{%
}{.}\hspace{.4pt}8}}


\bibitem{sayes2014actor}
E.~Sayes.
\newblock Actor--network theory and methodology: Just what does it mean to say that nonhumans have agency?
\newblock {\em Social studies of science}, 44(1):134--149, 2014. \href{https://doi.org/10.1177/0306312713511867}
{doi: \textsf{%
10\hspace{.1pt}\discretionary{.}{%
}{.}\hspace{.4pt}1177\discretionary{/}{%
}{/}0306312713511867}}


\bibitem{schieffelin_language_1986}
B.~B. Schieffelin and E.~Ochs.
\newblock Language {Socialization}.
\newblock {\em Annual Review of Anthropology}, 15:163--191, Oct. 1986.
\newblock Publisher: Annual Reviews. \href{https://doi.org/10.1146/annurev.an.15.100186.001115}
{doi: \textsf{%
10\hspace{.1pt}\discretionary{.}{%
}{.}\hspace{.4pt}1146\discretionary{/}{%
}{/}annurev\hspace{.1pt}\discretionary{.}{%
}{.}\hspace{.4pt}an\hspace{.1pt}\discretionary{.}{%
}{.}\hspace{.4pt}15\hspace{.1pt}\discretionary{.}{%
}{.}\hspace{.4pt}100186\hspace{.1pt}\discretionary{.}{%
}{.}\hspace{.4pt}001115}}


\bibitem{schofieldIndexicalityVisualizationExploring2013a}
T.~Schofield, M.~D{\"o}rk, and M.~{Dade-Robertson}.
\newblock Indexicality and visualization: Exploring analogies with art, cinema and photography.
\newblock In {\em Proceedings of the 9th {{ACM Conference}} on {{Creativity}} \& {{Cognition}}}, pp. 175--184. ACM, New York, USA, June 2013. \href{https://doi.org/10.1145/2466627.2466641}
{doi: \textsf{%
10\hspace{.1pt}\discretionary{.}{%
}{.}\hspace{.4pt}1145\discretionary{/}{%
}{/}2466627\hspace{.1pt}\discretionary{.}{%
}{.}\hspace{.4pt}2466641}}


\bibitem{segelNarrativeVisualizationTelling2010}
E.~Segel and J.~Heer.
\newblock Narrative {Visualization}: {Telling} {Stories} with {Data}.
\newblock {\em IEEE Transactions on Visualization and Computer Graphics}, 16(6):1139--1148, Nov. 2010. \href{https://doi.org/10.1109/TVCG.2010.179}
{doi: \textsf{%
10\hspace{.1pt}\discretionary{.}{%
}{.}\hspace{.4pt}1109\discretionary{/}{%
}{/}TVCG\hspace{.1pt}\discretionary{.}{%
}{.}\hspace{.4pt}2010\hspace{.1pt}\discretionary{.}{%
}{.}\hspace{.4pt}179}}


\bibitem{silverstein1979ideology}
M.~Silverstein.
\newblock Language structure and linguistic ideology.
\newblock In {\em The Elements: A parasession of linguistic units and levels.} Chicago Linguistic Society, Chicago, IL, US, 1979.

\bibitem{small_how_2009}
M.~L. Small.
\newblock `{How} many cases do {I} need?': {On} science and the logic of case selection in field-based research.
\newblock {\em Ethnography}, 10(1):5--38, Mar. 2009.
\newblock Publisher: SAGE Publications. \href{https://doi.org/10.1177/1466138108099586}
{doi: \textsf{%
10\hspace{.1pt}\discretionary{.}{%
}{.}\hspace{.4pt}1177\discretionary{/}{%
}{/}1466138108099586}}


\bibitem{tiidenberg_tumblr_2021}
K.~Tiidenberg, N.~Hendry, and C.~Abidin.
\newblock {\em Tumblr}.
\newblock Polity Press, Cambridge, UK, 2021.

\bibitem{Tufte1985}
{\relax ER}.~Tufte.
\newblock {\em Visual {{Display}} of {{Quantitative Information}}}.
\newblock Graphics Paper Press LLC, Cheshire, CT, 1983.

\bibitem{tverskyCognitivePrinciplesGraphic1997}
B.~Tversky.
\newblock Cognitive {{Principles}} of {{Graphic Displays}}.
\newblock In {\em Proceedings from the AAAI Fall Symposium}, 1997.

\bibitem{Ware2004}
C.~Ware.
\newblock {\em Information {Visualization}: {Perception} for {Design}}.
\newblock Elsevier Inc., San Francisco, CA, 2004.

\bibitem{woolardCodeswitching2004}
K.~A. Woolard.
\newblock Codeswitching.
\newblock In {\em A {Companion} to {Linguistic} {Anthropology}}. Wiley \& Sons, 2004. \href{https://doi.org/10.1002/9780470996522.ch4}
{doi: \textsf{%
10\hspace{.1pt}\discretionary{.}{%
}{.}\hspace{.4pt}1002\discretionary{/}{%
}{/}9780470996522\hspace{.1pt}\discretionary{.}{%
}{.}\hspace{.4pt}ch4}}


\bibitem{woolard1990changing}
K.~A. Woolard and T.-J. Gahng.
\newblock Changing language policies and attitudes in autonomous catalonia.
\newblock {\em Language in Society}, 19(3):311--330, 1990. \href{https://doi.org/10.1017/S0047404500014536}
{doi: \textsf{%
10\hspace{.1pt}\discretionary{.}{%
}{.}\hspace{.4pt}1017\discretionary{/}{%
}{/}S0047404500014536}}


\bibitem{wutichSampleSizes102024}
A.~Wutich, M.~Beresford, and H.~R. Bernard.
\newblock Sample {{Sizes}} for 10 {{Types}} of {{Qualitative Data Analysis}}: {{An Integrative Review}}, {{Empirical Guidance}}, and {{Next Steps}}.
\newblock {\em International Journal of Qualitative Methods}, 23, Jan. 2024. \href{https://doi.org/10.1177/16094069241296206}
{doi: \textsf{%
10\hspace{.1pt}\discretionary{.}{%
}{.}\hspace{.4pt}1177\discretionary{/}{%
}{/}16094069241296206}}


\bibitem{yin_case_2002}
R.~Yin, K.
\newblock {\em Case study research: {Design} and methods}.
\newblock Sage, 2002.

\end{thebibliography}
\end{document}